# *Proving Correctness and Completeness of Normal Programs — a Declarative Approach*

WŁODZIMIERZ DRABENT

*Institute of Computer Science, Polish Academy of Sciences,*
*ul. Ordona 21, Pl – 01-237 Warszawa, Poland*
*and*
*Linköpings universitet, Department of Computer and Information Science*
*S – 581 83 Linköping, Sweden*
(*e-mail:* `drabent@ipipan.waw.pl`)

MIROSŁAWA MIŁKOWSKA

*Institute of Informatics, Warsaw University, ul. Banacha 2, 02-097 Warszawa, Poland*
(*e-mail:* `M.Milkowska@mimuw.edu.pl`)

## Abstract

We advocate a declarative approach to proving properties of logic programs. Total correctness can be separated into correctness, completeness and clean termination; the latter includes non-floundering. Only clean termination depends on the operational semantics, in particular on the selection rule. We show how to deal with correctness and completeness in a declarative way, treating programs only from the logical point of view. Specifications used in this approach are interpretations (or theories). We point out that specifications for correctness may differ from those for completeness, as usually there are answers which are neither considered erroneous nor required to be computed.

We present proof methods for correctness and completeness for definite programs and generalize them to normal programs. For normal programs we use the 3-valued completion semantics; this is a standard semantics corresponding to negation as finite failure. The proof methods employ solely the classical 2-valued logic. We use a 2-valued characterization of the 3-valued completion semantics, which may be of separate interest.

The method of proving correctness of definite programs is not new and can be traced back to the work of Clark in 1979. However a more complicated approach using operational semantics was proposed by some authors. We show that it is not stronger than the declarative one, as far as properties of program answers are concerned. For a corresponding operational approach to normal programs, we show that it is (strictly) weaker than our method. We also employ the ideas of this work to generalize a known method of proving termination of normal programs.

## 1 Introduction

This paper discusses reasoning about logic programs in terms of their declarative semantics. We view total correctness of programs as consisting of correctness, com-pleteness and clean termination. *Correctness* (sometimes called partial correctness) means that any answer obtained from the program satisfies the specification. As logic programming is nondeterministic, one is interested in *completeness*, i.e. that all the results required by the specification are computed. Programs should also (cleanly) *terminate* — computations should be finite and without run-time errors, like floundering and arithmetical exceptions.

Obviously, clean termination depends on the operational semantics, in particular on the selection rule. However correctness and completeness do not; they are *declarative properties*. It is desirable that they could be dealt with in a declarative way, abstracting from any operational semantics and treating programs and their answers only from the logical point of view. Otherwise logic programming would not deserve to be considered a declarative programming paradigm. Declarative treatment of correctness and completeness makes it possible to separate reasoning about "logic" and "control"; correctness and completeness are related to logic and clean termination to control. Changing the control component does not influence correctness and completeness.

In this paper we show how to prove correctness and completeness declaratively. We discuss a known method of proving correctness of definite programs and introduce a method for proving completeness. Then we generalize both methods to programs with negation. As their declarative semantics we employ the 3-valued completion semantics (Kunen 1987). Our proof methods use however only the standard 2-valued logic. The employed 2-valued characterization of Kunen semantics may be of separate interest.

The proof method for definite program correctness (Clark 1979; Hogger 1981; Deransart 1993) is simple and straightforward. It is declarative: it abstracts from any operational semantics. It should be well known. However its usefulness is often not appreciated. Instead a more complicated approach using operational semantics was proposed by some authors (Bossi and Cocco 1989; Apt 1997; Pedreschi and Ruggieri 1999). That approach takes into account the form of atoms selected under LD-resolution. We show that, as far as declarative properties of programs are concerned, the operational approach is not stronger than the declarative one. The last of these papers also deals with normal programs. In this case we show that the operational approach is strictly weaker than that presented here, when declarative properties are of interest.

The following observation is important for our approach: it should be possible to use approximate specifications, and one should not require that the same specification is used for both correctness and completeness. This is natural, as there usually are answers which are neither considered erroneous nor required to be computed. Using the same specification for both purposes requires making decisions like "should $append([\,],7,7)$ be correct?"; this brings substantial and unnecessary complications. So there is some 3-valued flavour even in logic programming without negation. Notice that if a program is both correct and complete with respect to a specification then the specification cannot be approximate. Approximate specifications are useful not only in the context of proving correctness and completeness. We show how a (non-unique) approximate specification can replace the unique in-terpretation in the method of (Apt and Pedreschi 1993) for proving termination of normal programs.

The paper consists of two main chapters: Section 3 is devoted to definite programs, Section 4 to normal programs. In each case we first discuss proving correctness, then proving completeness. We also discuss completeness of the presented proof methods and compare them with the operational approach. Section 4.3 on proving correctness of normal programs also presents a generalization of the method for proving termination by Apt and Pedreschi (1993). The paper is concluded by a section on related work. A preliminary and abridged version of this paper appeared as (Drabent and Miłkowska 2001).

## 2 Preliminaries

For basic definitions we refer the reader to (Lloyd 1987) and to (Apt 1997; Doets 1994). We consider the declarative semantics given by 3-valued logical consequences of program completion (Kunen 1987). This is a standard semantics for normal programs with finite failure (Doets 1994). It is a generalization of the classical semantics for definite programs (2-valued logical consequences of the program). SLDNF-resolution is sound for this semantics and important completeness results exist.

We are interested in declarative properties of programs, i.e. properties of programs treated as sets of logic formulae. Speaking more formally, we consider properties of program answers. We are not interested in distinguishing logically equivalent programs, for instance logically equivalent definite programs with different S-semantics (Bossi et al. 1994), like $\{\,p(X)\!\leftarrow,\ p(a)\!\leftarrow\,\}$ and $\{\,p(X)\!\leftarrow\,\}$.

By a computed (resp. correct) answer we mean an instance $Q\theta$ of a query $Q$, where $\theta$ is a computed (correct) answer substitution for $Q$ and the given program. (A query is a sequence of literals; it is a sequence of atoms when definite programs are concerned). Notice that, by soundness and completeness of SLD-resolution, the sets of computed and of correct answers for a given definite program are equal. (In particular, a correct answer $Q\theta$ for a query $Q$ is a computed answer for a query $Q\theta$.) So in the case of definite programs we usually do not distinguish between these two kinds of answers; the term "answer" refers to both of them. Due to incompleteness of SLDNF-resolution, some correct answers for normal programs are not computed answers. So in the context of normal programs the term "answer" refers to correct answers.

We assume an arbitrary fixed first order language $\mathcal{L}$. Sometimes it is required that the set of function symbols of $\mathcal{L}$ is infinite; this will be stated explicitly. A *preinterpretation* for $\mathcal{L}$ is an algebra $\mathcal{J}$ and a mapping assigning an $n$-ary function of $\mathcal{J}$ to each $n$-ary ($n \geq 0$) function symbol of $\mathcal{L}$. We will represent interpretations as sets (Lloyd 1987, p. 12), (Doets 1994, p. 124): an interpretation (over $\mathcal{J}$) is a set of constructs of the form $p(e_1, \ldots, e_n)$, where $p$ is a predicate symbol and $e_1, \ldots, e_n$ are elements of the carrier $|\mathcal{J}|$ of $\mathcal{J}$. Such a $p(e_1, \ldots, e_n)$ will be called a $\mathcal{J}$-atom. Obviously, if $\mathcal{J}$ is a Herbrand algebra then an interpretation is a set of ground atoms. The Herbrand base w.r.t. (with respect to) a given language $\mathcal{L}$ will be denoted by $\mathcal{H}$, and the least Herbrand model of a definite program $P$ by $M_P$.

We sometimes use a comma instead of $\wedge$ and assume that conjunction has a higher priority than disjunction, and disjunction higher than implication. For instance $\alpha, \beta \vee \gamma, \delta$ stands for $(\alpha \wedge \beta) \vee (\gamma \wedge \delta)$, and $\alpha \vee \beta \to \gamma$ for $(\alpha \vee \beta) \to \gamma$. In program examples we use some elements of the notation of Prolog (variable names begin with an upper case letter, lists are denoted using $[,|,]$, etc).

## 3  Reasoning about Definite Programs

First we show a method of proving program correctness. In the next section we compare it with an approach related to operational semantics. Then we introduce a method of proving completeness.

### 3.1  Correctness of Definite Programs

We begin with a brief discussion on specifications. As a standard example let us take the program APPEND:

$$app(\,[\,], L, L\,) \leftarrow$$
$$app(\,[H|K], L, [H|M]\,) \leftarrow app(\,K, L, M\,)$$

We want to prove that it indeed appends lists. We need a precise statement (a specification) of this property. A slight complication is that the program does not actually define the relation of list concatenation, but its superset; the least Herbrand model contains atoms like $app([\,], 1, 1)$. This is a common phenomenon in logic programming, the least model contains "ill-typed" atoms which are irrelevant for the correctness of the program.

So we want to prove that:

for any answer $app(k, l, m)$, if $k$ and $l$ are lists then $m$ is a list and $k * l = m$.

(By a list we mean a term $[t_1, \ldots, t_n]$ (in Prolog notation), where $n \geq 0$ and $t_1, \ldots, t_n$ are possibly non-ground terms. Symbol $*$ denotes the list concatenation.) This property could be equivalently expressed as

$$spec \models app(k, l, m) \qquad (1)$$

for any answer $app(k, l, m)$, where $spec$ is the Herbrand interpretation:

$$spec = \{\, app(k, l, m) \in \mathcal{H} \mid \text{if } k \text{ and } l \text{ are lists then } m \text{ is a list and } k * l = m\,\} \quad (2)$$

Obviously, (1) holds iff all the ground instances of $app(k, l, m)$ are in $spec$.

Notice that we do not need to refer to the notion of a query in the specification. Assume that $app(k, l, m) = app(k', l', m')\theta$ is a computed answer for a query $app(k', l', m')$. If $k', l'$ are lists then obviously $k, l$ are lists and (1) implies that $m$ is a list and $k * l = m$.

Such specifications, referring to program answers, will be called *declarative*. A declarative specification can be an interpretation (possibly not a Herbrand one) or a theory.[1] In this paper we will use specifications of the first kind, but we expect that our results also apply to specifications of the second kind.

*Definition 3.1*
A definite program is **correct** w.r.t. a declarative specification *spec* iff $spec \models Q$ for any answer $Q$ of the program.

Notice that a program $P$ is correct w.r.t. a Herbrand interpretation *spec* iff its least Herbrand model $M_P$ is a subset of *spec* (as for such interpretations $spec \models Q$ means that all the ground instances of the atoms in $Q$ are in *spec*).

To prove correctness (of a logic program w.r.t. a declarative specification) we use an obvious approach, discussed among others by Clark (1979), Hogger (1981, p. 378–9) and Deransart (1993, Section 3).[2] We will call it the *natural* proof method. It consists of showing that $spec \models C$ for each clause $C$ of the considered program. The soundness of the natural method follows from the following simple property.

*Proposition 3.2* (*Correctness, definite programs*)
Let $P$ be a program and *spec* be an interpretation. If

$$spec \models P$$

then $P$ is correct w.r.t. specification *spec*.

*Proof*
By soundness of SLD-resolution, $P \models Q$ for any answer $Q$. Now $spec \models P$ and $P \models Q$ imply $spec \models Q$. (This also holds for *spec* being a theory.) □

The method is also complete (Deransart 1993) in the following sense. If a program $P$ is correct w.r.t. a declarative specification *spec* then there exists a stronger specification $spec' \subseteq spec$ such that $spec' \models P$, and thus the method is applicable to $spec'$. (To prove this property, take as $spec'$ the least model of $P$ over the given preinterpretation.)

*Example 3.3*
The proof of correctness of APPEND w.r.t. specification (2) is rather simple. We present here its less trivial part with details. Consider the second clause. To show that

$$spec \models app([H|K], L, [H|M]) \leftarrow app(K, L, M)$$

take ground terms $h, k, l, m$ (and valuation $\{H/h, K/k, L/l, M/m\}$) such that $spec \models app(k, l, m)$, in other words $app(k, l, m) \in spec$. We have to show that $spec \models app([h|k], l, [h|m])$. Assume that $[h|k]$ and $l$ are lists, hence $k$ is a list. Then $m$ is a list and $k * l = m$, as $spec \models app(k, l, m)$. Thus $[h|m]$ is a list and $[h|k] * l = [h|m]$, hence $app([h|k], l, [h|m]) \in spec$. This concludes the proof.

---

[1] A specification corresponding to our example specification *spec* may consist of an axiom $app(k, l, m) \leftrightarrow (list(k), list(l) \rightarrow list(m), k{*}l{=}m)$ together with axioms describing predicates $=$, *list* and function $*$, and an induction schema for lists.
[2] where it is called "inductive proof method".

*Example 3.4*
The specification of APPEND considered above does not describe the usage of APPEND to split a list or to subtract lists. Also the requirement on $k$ is unnecessary. This is because our intention was to follow a corresponding example of (Apt 1997). A full specification of APPEND may be

$$spec_{\mathrm{APPEND}} = \left\{\, app(k,l,m) \in \mathcal{H} \;\middle|\; \begin{array}{l} \text{if } l \text{ or } m \text{ is a list then} \\ k,l,m \text{ are lists and } k*l = m \end{array} \,\right\}.$$

It is easy to check, in a way described above, that $spec_{\mathrm{APPEND}} \models \mathrm{APPEND}$. Thus by Proposition 3.2 program APPEND is correct w.r.t. $spec_{\mathrm{APPEND}}$.

The program in the next example uses accumulators. Alternatively it can be seen as employing difference lists. Let us define that a difference list representing a list $[t_1,\ldots,t_n]$ is any pair $([t_1,\ldots,t_n|t], t)$ of terms, where $t$ is an arbitrary term.

*Example 3.5*
Consider the standard REVERSE program:

$$\begin{aligned} &reverse(X,Y) \leftarrow rev(X,Y,[\,]) \\ &rev([\,],X,X) \leftarrow \\ &rev([H|L],X,Y) \leftarrow rev(L,X,[H|Y]) \end{aligned}$$

The declarative reading of the program is simple: the first argument of *rev* is a list, its reverse is represented as a difference list of the second and the third argument. This can be expressed by a formal specification

$$\begin{aligned} spec_{\mathrm{R}} \;\; &= \;\; \{\, reverse([t_1,\ldots,t_n],[t_n,\ldots,t_1]) \mid n \geq 0,\; t_1,\ldots,t_n \in \mathcal{T} \,\} \\ &\cup \;\; \{\, rev([t_1,\ldots,t_n],[t_n,\ldots,t_1|t],t) \mid n \geq 0,\; t_1,\ldots,t_n,t \in \mathcal{T} \,\} \end{aligned}$$

where $\mathcal{T}$ is the set of ground terms.

To prove that the program is correct w.r.t. this specification it is sufficient to show $spec_{\mathrm{R}} \models \mathrm{REVERSE}$. The nontrivial part of the proof is to show that the last clause is true in the interpretation $spec_{\mathrm{R}}$. Take ground terms $l,x,h,y$, such that $spec_{\mathrm{R}} \models rev(l,x,[h|y])$. So there exist $n \geq 0$, $t_1,\ldots,t_n,t$ such that $l = [t_1,\ldots,t_n]$, $x = [t_n,\ldots,t_1|t]$, $t = [h|y]$. Then $rev([h|l],x,y)$ is $rev([h,t_1,\ldots,t_n],[t_n,\ldots,t_1,h|y],y)$, thus $spec_{\mathrm{R}} \models rev([h|l],x,y)$.

A quite common opinion is that "ill-typed" logical consequences of programs (like $app([\,],1,1)$ for program APPEND) lead to difficulties in reasoning about program correctness (cf. eg. (Apt 1995; Apt 1997; Naish 1992)). Similarly, programs dealing with accumulators or difference lists are sometimes considered difficult to reason about(cf. eg. (Apt 1995)). The natural method deals with such programs without any special burden, as the examples above show.

Notice that the natural method refers only to the declarative semantics of programs. A specification is an interpretation (alternatively a theory). Correctness is expressed as truth (of the program's answers) in the interpretation. Program clauses are treated as logic formulae, their truth in the interpretation is to be shown. We abstract from any operational semantics, in particular from the form of queries appearing during computation. The reasoning is obviously independent from the selection rule. Still we can use declarative specifications to reason about queries and corresponding answers, using the fact that an answer is an instance of the query.

### *3.2 Call-Success Specifications and the Operational Approach*

In this section we present an operational approach to program correctness and prove that it is not stronger than the natural method of Proposition 3.2 (as far as properties of program answers are concerned). We also argue that from a practical point of view the natural method is advantageous.

Some authors (Bossi and Cocco 1989), (Apt 1997, Chapter 8), (Pedreschi and Ruggieri 1999)[3] propose another approach to proving correctness. The approach explicitly deals with the form of queries. It uses specifications consisting of two parts. The **precondition** specifies atomic queries and the **postcondition** their success instances. We will call such specifications **call-success specifications**. Formally, pre- and postconditions are sets of atoms, closed under substitutions.

The proof method used in this approach was proposed by (Bossi and Cocco 1989) and is an instance of the method of (Drabent and Małuszyński 1988).[4] We will call it the *operational* proof method. It is based on the following verification condition:

Let $\langle pre, post \rangle$ be a call-success specification, with the precondition *pre* and the postcondition *post*. For each clause $C$ of the program it should be shown that for each (possibly non-ground) instance $H \leftarrow B_1, \ldots, B_n$ $(n \geq 0)$ of $C$

$$\begin{aligned}&\text{if } H \in pre, \ B_1, \ldots, B_k \in post \text{ then } B_{k+1} \in pre \text{ (for } k=0,\ldots,n{-}1\text{)},\\&\text{if } H \in pre, \ B_1, \ldots, B_n \in post \text{ then } H \in post.\end{aligned} \quad (3)$$

Additionally, there is a condition on initial queries. One requires that for any instance $B_1, \ldots, B_n$ $(n > 0)$ of such query, if $B_1, \ldots, B_k \in post$ then $B_{k+1} \in pre$ (for $k = 0, \ldots, n{-}1$). In (Apt 1997) a program (a query) with a call-success specification is called well-asserted if it satisfies the respective condition above.

The intuition behind condition (3) is related to operational semantics – to procedure calls and successes under LD-resolution (SLD-resolution with the Prolog selection rule). Indeed, (3) implies a stronger, non-declarative notion of correctness. We will say that a program is **correct** *w.r.t. a call-success specification* $\langle pre, post \rangle$ if every procedure call in an LD-derivation is in *pre* and every procedure success is in *post*, provided the initial query satisfies the condition above. By a procedure call we mean the atom selected in a goal, and by a procedure success a computed instance of a procedure call. If $P$ satisfies the verification condition (3) then $P$ is correct w.r.t. the call-success specification (see (Apt 1997) and the references therein).

Notice that correctness w.r.t. a call-success specification is not a declarative property. It considers not only computed answers, but whole computations (LD-trees). Thus this kind of correctness depends on the selection rule used. This is why we call the method operational.

---
[3] Whenever these approaches differ, we follow that of (Apt 1997).
[4] The latter approach does not require specifications to be closed under substitutions.

*Example 3.6*
Consider the APPEND program. We refer here to its treatment in (Apt 1997, p. 214). The precondition and postcondition are, respectively,

$$pre = \{\, app(k,l,m) \mid k \text{ and } l \text{ are lists}\,\},$$
$$post = \{\, app(k,l,m) \mid k,l,m \text{ are lists and } k*l = m \,\}.$$

(Here $k, l, m$ are terms, possibly non-ground.) The details of the proof can be found in (Apt 1997).

Now we formally compare both proof methods. We are going to prove that, as far as declarative properties are of interest, both methods are equivalent. Remember that we refer to two notions of program correctness: w.r.t. declarative specifications (of the natural method) and w.r.t. call-success specifications (of the operational method).

We first prove that the operational method is stronger than the natural one. We show that correctness w.r.t. a declarative specification can be expressed by means of correctness w.r.t. a call-success specification, and that whatever can be proven by the natural method, can be proven by the operational method. Roughly speaking, the natural method is the operational one with the preconditions abandoned.

*Proposition 3.7*
Let $P$ be a program, and let an interpretation *spec* be a declarative specification. Consider a call-success specification $\langle pre^\top, post(spec) \rangle$, where $pre^\top$ is the set of all atoms and $post(spec) = \{A \mid spec \models A\}$.

Then $P$ is correct w.r.t. *spec* iff $P$ is correct w.r.t. $\langle pre^\top, post(spec) \rangle$. Moreover, $P$ and *spec* satisfy the verification condition of the natural method (Proposition 3.2) iff $P$ and $\langle pre^\top, post(spec) \rangle$ satisfy the verification condition (3) of the operational method.

*Proof*
The first equivalence is obvious.

Consider the call-success specification $\langle pre^\top, post(spec) \rangle$. Notice that the condition on initial queries is trivially satisfied by any query. All the implications of (3) except the last one are trivially satisfied. The last implication of (3) holds for each instance of a clause $C$ iff $spec \models C$. (For the non-obvious "if" case notice that $B_1, \ldots, B_n \in post(spec)$ means $spec \models B_i$ for $i = 1, \ldots, n$; hence from $spec \models H \leftarrow B_1, \ldots, B_n$ we obtain $spec \models H$.) □

It remains to show that the operational method is not stronger than the natural one, as far as the declarative properties are concerned. Consider a call-success specification $\langle pre, post \rangle$. A corresponding declarative specification could be seen, speaking informally, as implication $pre \to post$. The following definition formalizes this idea.

*Definition 3.8*

Let *pre* and *post* be sets of atoms closed under substitution. The *declarative specification corresponding* to the call-success specification $\langle pre, post \rangle$ is the Herbrand interpretation

$$pre \rightarrow post \ := \ \{\, A \in \mathcal{H} \mid \text{if } A \in pre \text{ then } A \in post \,\}.$$

In other words, $pre \rightarrow post = (\mathcal{H} \setminus pre) \cup (\mathcal{H} \cap post)$. If $P$ is correct w.r.t. $pre \rightarrow post$ and $A\theta$ is an answer to a query $A \in pre$ then $A\theta \in post$. As an example take the call-success specification of APPEND from Example 3.6. The corresponding declarative specification is the specification (2) of APPEND from the previous section.

The following proposition compares the corresponding declarative and call-success specifications. (Similar property is mentioned without proof in (de Boer et al. 1997; Pedreschi and Ruggieri 1999).) The next proposition (see also (Courcelle and Deransart 1988)) compares both proof methods.

*Proposition 3.9*
If a program $P$ is correct w.r.t. a call-success specification $\langle pre, post \rangle$ then $P$ is correct w.r.t. the declarative specification $pre \rightarrow post$.

*Proof*
Assume that a program $P$ is correct w.r.t. $\langle pre, post \rangle$. As $pre \rightarrow post$ is a Herbrand interpretation, it is sufficient to show that $M_P \subseteq pre \rightarrow post$ (cf. the comment following Definition 3.1). Consider an $A \in M_P$. So query $A$ succeeds. If $A \in pre$ then $A \in post$. Thus $A \in pre \rightarrow post$. □

Now we show that if it can be proved by the operational method that a program $P$ is correct w.r.t. $\langle pre, post \rangle$ then it can be proved by the natural method that $P$ is correct w.r.t. $pre \rightarrow post$.

*Proposition 3.10*
If $P$ and $\langle pre, post \rangle$ satisfy the verification condition (3) of the operational method then $pre \rightarrow post \models P$.

*Proof*
$pre \rightarrow post \models P$ means that for any ground instance $H \leftarrow B_1, \ldots, B_n$ of a clause of $P$, if $B_1, \ldots, B_n \in pre \rightarrow post$ then $H \in pre \rightarrow post$. Consider such an instance and assume that $B_1, \ldots, B_n \in pre \rightarrow post$. If $H \notin pre$ then $H \in pre \rightarrow post$. Otherwise, for $H \in pre$ we obtain from (3) by simple induction that $B_i \in pre$ and $B_i \in post$, for $i = 1, \ldots, n$. Hence $H \in post$, by (3); thus $H \in pre \rightarrow post$. □

The converse of the two propositions does not hold:

*Example 3.11*
For a simple counterexample consider $P$ and $\langle pre, post \rangle$ satisfying the verification condition (3), and reorder the atoms in the clause bodies. The obtained program $P_1$ may be incorrect w.r.t. $\langle pre, post \rangle$, but $pre \rightarrow post \models P_1$.

For a counterexample independent from the ordering of body atoms, consider the program $P_2$:

$$p(X,Y,Z) \leftarrow q(X,Z), q(Y,Z)$$
$$q(X,X) \leftarrow$$

Let $S$ be a set of ground terms. Consider a declarative specification *pre→post*, where

$$\begin{aligned} pre &= \{p(t_1,t_2,t_3) \mid t_1 \in S \text{ or } t_2 \in S\} \cup \{q(t_1,t_2) \mid t_1 \in S\}, \\ post &= \{p(t_1,t_2,t_3) \mid t_3 \in S\} \cup \{q(t_1,t_2) \mid t_2 \in S\}. \end{aligned}$$

We have *pre→post* $\models P_2$, but $P_2$ is incorrect w.r.t. the operational specification $\langle pre, post \rangle$. The same holds for $P_2$ with the atoms in the clause body swapped.

The last two propositions show that the natural method is stronger than the operational one (and hence equivalent to it), as far as declarative properties are concerned. In contrast to the operational method, the natural one is independent from the order of the body atoms in clauses.

We proved that the two methods are formally equivalent. Now we argue that, from the practical point of view, switching from the operational method to the natural one does not bring any difficulties or complications. First, a declarative specification corresponding to a given call-success specification is obtained from the latter by a simple composition of three operations: removing non-ground atoms, set complementation and set union. Then, the proof of Proposition 3.10 shows how to obtain a natural method proof out of an operational one. This is done by adding a few simple steps. (Notice that for the new proof we consider implications (3) only for ground instances of clauses).

The natural method requires proving one implication per program clause. In contrast, the operational method requires proving one implication for each atom occurring in the program or in the initial query. This is a price for obtaining more information: the property proved concerns not only program answers but also calls and successes under LD-derivations. However, when one is not interested in the latter, the natural method seems more convenient.

There are many examples of using the operational method where the declarative one is sufficient (for instance cf. the papers mentioned above). Apparently people are often confused by the fact that the least Herbrand model contains undesired, "ill-typed" atoms (cf. the opinions of (Apt 1995)). They want a specification describing exactly the set of atoms of interest. For instance, such a set for APPEND is $spec' = \{\, app(k,l,m) \mid k,l,m \text{ are lists and } k*l = m \,\}$. A program is usually not correct w.r.t. such a declarative specification. It is often not recognized that the property of interest can be described by means of an approximate declarative specification, like those from the examples of Section 3.1.

There may be another reason for using specifications describing exactly the sets of answers of interest: such specifications can be employed in reasoning about program completeness. It is however not necessary to use the same specification for both correctness and completeness. As we argue in Section 3.3, it is quite convenient and natural to use separate specifications instead.

Notice the difference in the treatment of "ill-typed" atoms (like $app([],1,1)$ for APPEND) by the two approaches. In the natural method we can *include* such atoms in a specification.[5] For example, the declarative specification $spec_{\text{APPEND}}$ from Example 3.4 contains all ground atoms of the form $app(k,l,m)$ where $l$ and $m$ are not lists. In the operational method the "ill-typed" atoms are usually *excluded* from postconditions. A precondition states explicitly how the program should be called to avoid "ill-typed" answers. The call-success specification of APPEND discussed above is a typical example. So the postcondition of a call-success specification is $M_P \cap pre$ or its superset, while a declarative Herbrand specification is a superset of $M_P$ (for instance $M_P \cup (\mathcal{H} \setminus pre)$).

The following example shows that such treatment of "ill-typed" atoms by the operational method is impossible in some cases. It also shows that sometimes a non-trivial precondition cannot be used and the operational method boils down to the natural one; what conceptually is a precondition has to be expressed by a postcondition.

*Example 3.12*
Let us consider a program $\text{TWO} = \text{TWO}_p \cup \text{TWO}_q$, where $\text{TWO}_p$ is

$$p(X,Y) \leftarrow q(X,X2,X1,X3), q(X1,X3,X2,Y)$$

and $p$ does not occur in $\text{TWO}_q$. Assume that $\text{TWO}_q$ is correct w.r.t. declarative specification

$$spec_q = (pre_q^1 \to post_q^3) \cap (pre_q^2 \to post_q^4),$$

where

$$\begin{aligned}pre_q^1 &= \{\, q(t,s,u,v) \mid list(t)\,\} & post_q^3 &= \{\, q(t,s,u,v) \mid list(u)\,\} \\ pre_q^2 &= \{\, q(t,s,u,v) \mid list(s)\,\} & post_q^4 &= \{\, q(t,s,u,v) \mid list(v)\,\}\end{aligned}$$

and $list(t)$ stands for "$t$ is a list", for a possibly non-ground term $t$. (Thus $spec_q$ states that if the $i$-th argument of $q$ is a list then its argument $i+2$ is a list too, for $i=1,2$. Nothing more is known about $\text{TWO}_q$. Notice that $spec_q$ includes all atoms with the predicate symbol distinct from $q$.)

Program TWO is an abstraction of "two-pass" programs and of certain usages of difference lists. Some examples of such programs can be found e.g. in (Boye and Małuszyński 1997). Informally, its data flow can be described as follows. The value of $X1$ produced by the first call of $q$ is used by the second call. The value of $X2$ is produced by the latter and used by the former, which uses it to produce the value of $X3$. The value of $X3$ is used by the second call, which produces the value of $Y$.

By means of the natural method it is easy to show that, in an answer $p(t_1,t_2)$ of TWO, if $t_1$ is a list then $t_2$ is a list too. In order to describe this, let us use declarative specification

$$spec_{\text{TWO}} = (pre_p \to post_p) \cap spec_q$$

---

[5] Generally: a specification may permit any answer $A$ that is not an instance of any query for which the program is intended to be used.

where

$$pre_p = \{\, p(t,s) \mid list(t) \,\} \qquad post_p = \{\, p(t,s) \mid list(s) \,\}.$$

TWO is correct w.r.t. $spec_{\mathrm{TWO}}$, as $spec_{\mathrm{TWO}} \models \mathrm{TWO}_p$[6] and $\mathrm{TWO}_q$ is correct w.r.t. $spec_{\mathrm{TWO}}$ by our initial assumption.[7]

Correctness of TWO w.r.t. $spec$ implies that if $p$ is called with the first argument being a list and succeeds then the second argument is bound to a list. Now we discuss how this can be proved using the operational method. Let $\mathcal{A}_r$ denote the set of all atoms with the predicate symbol $r$.

To express this property one needs a call-success specification $\langle pre, post \rangle$ such that $pre \cap \mathcal{A}_p = pre_p$ and $post \cap \mathcal{A}_p = post_p$. Assume that the operational proof method is applicable to this specification, in other words that the verification conditions (3) hold. Hence in any LD-derivation any procedure call is in $pre$ and any procedure success is in $post$, provided the initial goal is in $pre$.

As explained previously, a usual way of using the operational method is such that "ill-typed" atoms are excluded from the postcondition. This is impossible for program TWO, as in the computations (i.e. LD-derivations) started from a goal $A \in pre_p$, predicate $q$ may succeed with its second and fourth arguments being not lists.

Notice that the precondition $pre$ has to permit any value of the second, third and fourth argument of $q$, as during the computation $q$ is invoked with these arguments being variables. Formally, $pre$ contains any atom of the form $q(t_1, t_2, t_3, t_4)$ where $t_1$ is a list. These atoms are in $pre_q^1$, but (some of them) are not in $pre_q^2$. Thus $pre_q^2$ cannot be used as a precondition for $q$ (more precisely, $pre \cap \mathcal{A}_q$ cannot be $pre_q^2$, or a subset of $pre_q^2$).

We could use $pre_q^1$ as a precondition for $q$ (formally $pre \cap \mathcal{A}_q$ could be $pre_q^1$) provided that $q$ did not occur in any clause body of $\mathrm{TWO}_q$. Otherwise we have to use the trivial precondition for $q$ (formally $pre \cap \mathcal{A}_q = \mathcal{A}_q$), as nothing is known about the procedure calls in $\mathrm{TWO}_q$.[8]

Hence for the last implication of (3) for $\mathrm{TWO}_p$ to hold, the postcondition has to express that if the $i$-th argument of $q$ is a list then its argument $i+2$ is a list, for

$$\begin{array}{ll}
H \in pre_p \quad \text{implies} \quad B_1 \in pre_q^1, & B_1 \in post_q^4 \quad \text{implies} \quad B_2 \in pre_q^2, \\
B_1 \in post_q^3 \quad \text{implies} \quad B_2 \in pre_q^1, & B_2 \in post_q^4 \quad \text{implies} \quad H \in post_p. \\
B_2 \in post_q^3 \quad \text{implies} \quad B_1 \in pre_q^2, &
\end{array}$$

---

[6] Here are the details of the proof. Take a ground instance $H \leftarrow B_1, B_2$ of the clause of $\mathrm{TWO}_p$. Notice that:

Assume that $spec_{\mathrm{TWO}} \models B_1, B_2$. Thus we have, for $i = 1, 2$:

$$B_i \in pre_q^1 \quad \text{implies} \quad B_i \in post_q^3, \qquad B_i \in pre_q^2 \quad \text{implies} \quad B_i \in post_q^4,$$

Combining these implications together we obtain that $H \in pre_p$ implies $H \in post_p$. This means that $spec_{\mathrm{TWO}} \models H$.

[7] To view this reasoning as application of Proposition 3.2, take a specification $I = M_{\mathrm{TWO}_q} \cup (pre_p \rightarrow post_p \cap \{p(t_1,t_2) \mid t_1, t_2 \text{ are ground terms}\})$. $I \subseteq spec_{\mathrm{TWO}}$ (as $M_{\mathrm{TWO}_q} \subseteq spec_q$), and $I \models \mathrm{TWO}$ (as $spec_{\mathrm{TWO}} \models \mathrm{TWO}_p$).

[8] Notice that the same holds if we swap the body atoms of $\mathrm{TWO}_p$.

$i = 1, 2$. So the postcondition for $q$ is the declarative specification $spec_q$ lifted to non-ground terms. Formally, $spec \cap \mathcal{A}_q = (\overline{pre_q^1} \cup post_q^3) \cap (\overline{pre_q^2} \cup post_q^4)$.

Conceptually, $pre_q^1$ and $pre_q^2$ are preconditions, as they are premises of implications which have to be used in the proof. However, as shown above, $pre_q^2$ (and $pre_q^1$, for some programs $\text{TWO}_q$) cannot be used in the precondition of the operational method. Instead they have to be employed in the postcondition.

Notice that the first two implications of (3) hold trivially, and that proving the last implication is basically a generalization of the declarative proof presented above.[9]

Thus the operational proof for $\text{TWO}_p$ is basically the same as the declarative method proof presented above. (Restriction of the former to ground clause instances gives the latter.)

Propositions 3.9 and 3.10 show that, for proving properties of program answers, we do not need preconditions. Declarative specifications and the natural method of Section 3.1 are sufficient. The proof of Proposition 3.10 shows how every operational method proof can be easily transformed into a natural method one, with introducing only minor changes. The converse does not hold; Examples 3.11 and 3.12 show that in some cases a natural method proof cannot be converted into an operational one with a non-trivial precondition.

The operational method is a generalization of the natural one: roughly speaking any natural method proof can be seen as operational one, with a trivial precondition (Proposition 3.7). The operational method proves more, as it also deals with the form of procedure calls and successes in LD-resolution. However it is more complicated and, for declarative properties, it is not stronger than the natural one. In our opinion, when one is interested only in declarative properties, the natural method should be preferred to the operational one.

### 3.3 Completeness of Definite Programs

Let us begin from an observation that for a given program a specification for completeness is in general different from that for correctness. For the purposes of correctness we describe a superset of the set of answers of a program. For the purposes of completeness we describe its subset, as a program satisfying a completeness requirement may compute something more than required. Often when a specification for correctness is of the form *pre→post* then a specification for completeness is *post*.

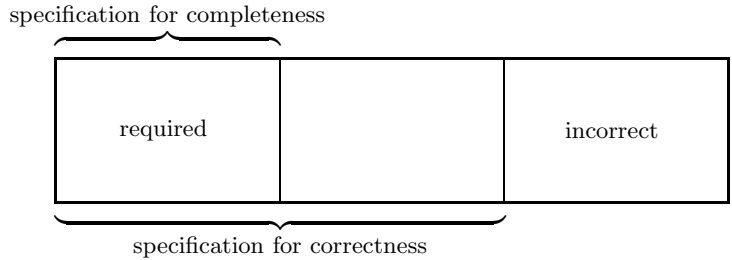

---

[9] One has to show that if $H \in pre_p$ and $B_1, B_2 \in (\overline{pre_q^1} \cup post_q^3) \cap (\overline{pre_q^2} \cup post_q^4)$ then $H \in post_p$, for any instance $H \leftarrow B_1, B_2$ of the clause of $\text{TWO}_p$. For ground instances this is equivalent to $spec_{\text{TWO}} \models \text{TWO}_p$.

For instance, it makes no sense to require that APPEND program were complete w.r.t. the specification *spec* from the beginning of Section 3.1, or $spec_{\text{APPEND}}$ from Example 3.4. Such a program should compute "ill-typed" answers, like $app(a,b,c)$. Our specification for completeness of APPEND is the Herbrand interpretation

$$specC_{\text{APPEND}} = \{\, app(k,l,m) \in \mathcal{H} \mid k,l,m \text{ are lists, } k*l = m \,\}.$$

Notice that it properly expresses our intentions: APPEND should compute all the cases of list concatenation. The difference $spec_{\text{APPEND}} \setminus specC_{\text{APPEND}}$ contains only "ill-typed" atoms, with the first or second argument not being a list. We are not interested whether they are answers of APPEND.

As previously, we consider specifications which are (possibly non-Herbrand) interpretations. Additionally we require that a specification is over a preinterpretation $\mathcal{J}$ in which equality satisfies the Clark equality theory CET.[10] (Alternatively, we may consider specifications which are theories containing CET.)

*Definition 3.13*
A definite program $P$ is **complete for a query** $Q$ w.r.t. a specification $specC$ if $specC \models Q\theta$ implies that $Q\theta$ is an answer for $P$, for any instance $Q\theta$ of $Q$.

$P$ is **complete** w.r.t. $specC$ if it is complete for any query w.r.t. $specC$.

Remember that $Q\theta$ is an answer for $P$ iff $P \models Q\theta$; this implies that $Q\theta$ is an instance of some computed answer for $Q$.

Below we refer to theory ONLY-IF($P$) (Doets 1994, p. 135) that is usually used while defining the Clark completion of a program $P$. Informally, ONLY-IF($P$) is $P$ with implications reversed. For each predicate symbol $p$, if the clauses of $P$ beginning with $p$ are $p(\vec{t}_1)\leftarrow\vec{B}_1,\ldots,p(\vec{t}_k)\leftarrow\vec{B}_k$ then ONLY-IF($P$) contains

$$p(\vec{x}) \;\to\; \bigvee_{i=1}^{k} \exists_{-\vec{x}}\; \vec{x} = \vec{t}_i \wedge \vec{B}_i,$$

where $\vec{x}$ are distinct new variables and the quantification is over the variables occurring in the clauses. For $k=0$ the implication is equivalent to $\neg p(\vec{x})$. In our example, ONLY-IF(APPEND) is (equivalent to)

$$app(x,y,z) \;\to\; x=[\,], y=z \;\vee\; \exists\, h,k,m\,(x=[h|k], z=[h|m], app(k,y,m)).$$

We also need a specification for equality:

$$spec_= = \{\, =(t,t) \mid t \in |\mathcal{J}|\,\}$$

where $|\mathcal{J}|$ is the carrier of the considered preinterpretation $\mathcal{J}$. So in the case of Herbrand specifications we have $spec_= = \{=(t,t) \mid t \text{ is a ground term}\}$.

---

[10] As an example what happens if this requirement is not satisfied consider $\mathcal{J}$ in which constants $a,b$ are given the same value. Take an interpretation *spec* over $\mathcal{J}$ such that $spec \models p(a)$. Then a program $P = \{\, p(a)\leftarrow\,\}$ is not complete w.r.t. *spec*, as $spec \models p(b)$ but $p(b)$ is not an answer of $P$.

The following can be used to prove completeness of a program.

*Proposition 3.14 (Completeness, definite programs)*
Let $P$ be a definite program and $Q$ a query. Assume that the set of function symbols of the underlying language is infinite. If

(i) $specC \cup spec_= \models \text{ONLY-IF}(P)$ and
(ii) $P$ terminates for $Q$, i.e. there exists a finite SLD-tree for $P$ and $Q$

then $P$ is complete for $Q$ w.r.t. $specC$.

If $P$ is complete w.r.t. $specC$ for each ground instance of each atom from a query $Q$ then $P$ is complete for $Q$ w.r.t. $specC$.

Notice that no particular selection rule is required in (ii). For Herbrand specifications, condition (i) means that for each $A \in specC$ there exists a ground instance $A \leftarrow B_1, \ldots, B_n$ of a clause of $P$ such that $B_1, \ldots, B_n \in specC$.

*Proof*
The first part of the proposition follows from a more general Theorem 4.21. (Take $spec = (\top, specC)$, where $\top$ is the set of all $\mathcal{J}$-atoms over the considered preinterpretation. By Theorem 4.6, $P$ seen as normal program is correct w.r.t. $spec$ and thus Theorem 4.21 can be applied.)

For the second part, let $Q = A_1, \ldots, A_n$ and $P$ be complete w.r.t. $specC$ for each ground instance of each $A_i$. Now $specC \models Q\theta$ implies $specC \models A_i\theta\sigma$ and then $P \models A_i\theta\sigma$, for each ground instance $A_i\theta\sigma$ of $A_i\theta$, $i = 1, \ldots, n$. This implies $P \models A_i\theta$. The latter follows from the fact that if $P \models A\sigma$ for each ground instance $A\sigma$ of an atom $A$ then $P \models A$ (see e.g. Theorem 3.3 of (Apt et al. 1996), the proof there requires infinite set of constants, but can be easily modified for infinite set of function symbols). We obtained $P \models A_i\theta$ for $i = 1, \ldots, n$, this means $P \models Q\theta$. □

*Example 3.15*
Consider program APPEND and the specification $specC_{\text{APPEND}}$ given above. It is easy to show that $specC_{\text{APPEND}} \cup spec_= \models \text{ONLY-IF}(\text{APPEND})$. One can show, using any standard method, that APPEND terminates for any ground atomic query. Thus APPEND is complete for any ground atomic query and then, by the second part of the proposition, complete.

Consider a query $Q = app(X, Y, m)$, where $X, Y$ are variables and $m$ a possibly non-ground list. For any lists $k, l$ such that $k * l = m$ we have $specC_{\text{APPEND}} \models app(k, l, m)$. By completeness of APPEND, $P \models app(k, l, m)$. So by completeness of SLD-resolution, $app(k, l, m)$ (or a more general atom) is a computed answer for $Q$. Summarizing, $Q$ produces all the required divisions of $m$ into two lists.

In our opinion, Proposition 3.14 is a formalization of a rather natural way of informal reasoning about completeness, which consists of checking that any tuple of argument values to be defined by the predicate is "covered" by some of its clauses.

The proposition without condition (ii) does not hold. Program $\{app(X, Y, Z) \leftarrow app(X, Y, Z)\}$ is a counterexample. Also the requirement on function symbols cannot be removed. For a counterexample take $P = \{p(a)\leftarrow, p(b)\leftarrow\}$, $Q = p(X)$, a Herbrand universe $\{a,b\}$ and a Herbrand interpretation $specC = \{\,p(a), p(b)\,\}$. Then $specC \models Q$ and $P$ is not complete for $Q$. However (i) and (ii) hold.

The method proposed here proves program completeness for queries that terminate. This should not be seen as a disadvantage, since in most cases termination of a program has to be established anyway. Deransart and Małuszyński (1993, Section 6.2.2, Theorem 6.1) provide a similar sufficient condition for completeness, which does not refer to termination. Instead of (ii) it employs some other property, which involves norms on atoms. Checking that property is similar to proving termination. So whenever termination has to be shown anyway, our approach is simpler.

Notice that Proposition 3.14 is also applicable when termination is not proven. Condition (i) alone implies that if we obtain a terminating execution for a particular query $Q$ then all the answers for $Q$ required by the specification have been computed.

There is certain limitation in using interpretations as specifications for completeness. One cannot express properties like "for any lists $k,l$ there exists some $m$ such that $P \models app(k,l,m)$." (The same limitation applies to call-success specifications of the operational approach from the previous section.) Such properties can however be expressed by specifications which are theories.

The proof method of Proposition 3.14 is not complete, in contrast to that of (Deransart and Małuszyński 1993). For a counterexample, consider a program $P$ containing a clause $p(\vec{x}) \leftarrow p(\vec{x})$. $P$ is complete (for query $p(\vec{x})$) w.r.t. specification $M_P$, but condition (ii) does not hold.

The method is however complete for arbitrary program $P$ and any query $Q$ for which $P$ terminates (or any query $A_1, \ldots, A_n$ such that $P$ terminates for any ground instance of $A_i$, for $i = 1, \ldots, n$). To prove this fact, assume that $P$ is complete w.r.t. a specification $specC$. Then there exists a weaker specification $specC' \supseteq specC$ such that $P$ is complete w.r.t. $specC'$ and (i) holds for $specC'$. One may take as $specC'$ the least model of $P$ over the given preinterpretation. Thus Proposition 3.14 makes it possible to show that, for any query as above, $P$ is complete w.r.t. $specC'$, and thus $specC$.

## 4 Reasoning about normal programs

We first discuss specifications for normal programs and present a 2-valued characterization of the 3-valued completion semantics. Then we introduce a method for proving correctness (Section 4.3) and a method for proving completeness of programs (Section 4.4). Each presentation includes an example, discussion of completeness of the method and comparison with an operational approach. In the section on correctness we also show how the presented approach can be used to generalize a well-known method for proving termination. We conclude with a bigger example (Section 4.5).

In this chapter, unless stated otherwise, the considered programs (queries) are normal programs (queries). We are interested in the completion semantics (logical consequences of program completion in 3-valued logic (Kunen 1987)). This semantics corresponds to an operational notion of finite failure. So we usually skip "finitely" in phrases like "finitely fails".

### *4.1 Specifications for normal programs*

In order to introduce specifications for normal programs let us first consider definite programs with queries which may contain negative literals. Assume that we have a definite program $P$ complete w.r.t. a Herbrand specification $specC$ and correct w.r.t. a $specS$ ($C$ as <u>c</u>ompleteness, $S$ as <u>s</u>oundness). If an atomic query $A$ fails then $specC \models \neg A$. So for $P$ and atomic queries, negation as finite failure is correct w.r.t. the specification for completeness. Now consider a query $Q = p(\vec{t}), \neg q(\vec{u})$. If it succeeds with an answer $Q\theta$ then $spec_1 \models Q\theta$ for an interpretation $spec_1$ that interprets $p$ as $specS$ and $q$ as $specC$. If $Q$ fails then $spec_2 \models \neg Q$ for an interpretation $spec_2$ that interprets $p$ as $specC$ and $q$ as $specS$. In order to deal with this phenomenon, we will use the following renamings of predicate symbols.

*Definition 4.1*
Let $\mathcal{L}$ be a first order language. Let $\mathcal{Q}$ be a formula or a set of formulae (e.g. a query or a program) of $\mathcal{L}$. Let us extend $\mathcal{L}$ by adding, for any predicate symbol $p$, a new predicate symbol $p'$.

$\mathcal{Q}'$ is $\mathcal{Q}$ with $p$ replaced by $p'$ in every negative literal of $\mathcal{Q}$ (for any predicate symbol $p$, except for =). Similarly, $\mathcal{Q}''$ is $\mathcal{Q}$ with $p$ replaced by $p'$ in every positive literal.

If $I$ is an interpretation for $\mathcal{L}$ then $I'$ is the interpretation obtained from $I$ by replacing each predicate symbol $p$ by $p'$.

For normal programs, a specification for correctness should describe two (possibly overlapping) sets of ground atoms — those allowed to succeed and those allowed to fail. Similarly, a specification for completeness should describe two (disjoint) sets, of the ground atoms required to fail and of those required to succeed. It is natural to allow to succeed any atom not required to fail, and allow to fail any atom not required to succeed. Hence the two sets needed to specify correctness can be the complements of the two sets used to specify completeness.

*Definition 4.2*
A **specification** for a normal program is a pair $(specS, specC)$, where $specC$ and $specS$ are interpretations over the same preinterpretation $\mathcal{J}$, in which the equality satisfies the Clark equality theory CET.

A specification $(specS, specC)$ is called **proper** if $specC \subseteq specS$.

Formal definitions of correctness and completeness are given in the respective sections below. For an informal explanation, assume that a program $P$ is correct w.r.t. a proper Herbrand specification $spec = (specS, specC)$. Then, if a ground atomic query $A$ succeeds then $A \in specS$, if it fails then $A \notin specC$. If $P$ is complete w.r.t. $spec$ then any $A \in specC$ succeeds and any $A \notin specS$ fails. Thus atomic queries from $specS - specC$ are allowed to succeed or to fail, but nothing is required about these queries.

One can consider correctness w.r.t. a specification that is not proper (atomic queries from $specC - specS$ are neither allowed to succeed nor to fail). On the other hand a program cannot be complete w.r.t. a non-proper Herbrand specification. This would require that some atoms both succeed and fail.

Sometimes it may be helpful to view our specifications as interpretations in the 4-valued logic of Belnap. The logical values in this logic are the subsets of $\{true, false\}$. By removing the value $\{true, false\}$ we obtain the logical values of the 3-valued logic that is usually used when dealing with semantics of logic programs. For a 4-valued interpretation $I$ and a formula $F$, $I \models_4 F$ means that the logical value of $F$ in $I$ contains $true$. For more details see (Fitting 1991) or (Stärk 1996). A specification $spec = (specS, specC)$ can be seen as a pair $I_S^4(spec), I_C^4(spec)$ of 4-valued interpretations. The first interpretation corresponds to viewing $spec$ as a specification for correctness, the other — for completeness.

*Definition 4.3*
Let $spec = (specS, specC)$ be a specification over a preinterpretation $\mathcal{J}$.

$I_S^4(spec)$ is the 4-valued interpretation over $\mathcal{J}$ such that for any $\mathcal{J}$-atom $A$ the logical value of $A$ contains $true$ iff $A \in specS$, and it contains $false$ iff $A \notin specC$.

$I_C^4(spec)$ is the 4-valued interpretation over $\mathcal{J}$ such that for any $\mathcal{J}$-atom $A$ the logical value of $A$ contains $true$ iff $A \in specC$, and it contains $false$ iff $A \notin specS$.

We will avoid 4-valued interpretations by employing the predicate renaming of Definition 4.1.

### 4.2 Characterization of 3-valued completion semantics

In this section we introduce a characterization of the 3-valued completion semantics of normal programs. The characterization uses the standard 2-valued logic. It employs renaming of predicate symbols. There exist other 2-valued characterizations of the completion semantics, based on predicate renaming. The approach of Mancarella et al. (1990) (see also references therein) is applicable to a restricted class of programs and deals with different semantics, which employs a domain closure axiom (thus the underlying language has a finite set of function symbols). Our characterization combines the ideas of those of (Stärk 1996) and (Drabent and Martelli 1991; Drabent 1996).

*Lemma 4.4 (Characterization of completion semantics)*
Let $P$ be a program and $Q$ a query.

$$comp(P) \models_3 Q \quad \text{iff} \quad P' \cup \text{ONLY-IF}(P'') \cup CET \models Q',$$
$$comp(P) \models_3 \neg Q \quad \text{iff} \quad P' \cup \text{ONLY-IF}(P'') \cup CET \models \neg Q''.$$

*Proof*
We use a result of Stärk (1996) who introduced a notion of partial completion of a logic program and showed that 3-valued consequences of the completion of a normal program are classical consequences of the partial completion of the program (modulo a simple syntactic transformations described below).

The main observation is that Stärk's partial completion $pcomp(P)$ of a program $P$ and $P' \cup \text{ONLY-IF}(P'') \cup CET$ are just syntactical transformations of each other.

Let $\mathcal{L}$ be the underlying first-order language. The language $\overline{\mathcal{L}}$ used by Stärk is obtained from $\mathcal{L}$ by adding for every predicate symbol $p$ a new symbol $\overline{p}$ with the same arity.

Let us transform $P'$, ONLY-IF($P''$), $Q'$ and $\neg Q''$ as follows:
– replace in ONLY-IF($P''$) each implication of the form $\alpha \to \beta$ by $\neg\beta \to \neg\alpha$,
– transform each formula containing negation to an equivalent form in which negation occurs only in negated literals,
– substitute each occurrence of a negated literal of the form $\neg p'(\vec{t})$ by atom $\overline{p}(\vec{t})$ (notice that every negated literal will be of that form and the obtained formulae do not contain primed predicate symbols).

Let us denote the translation of $F$ by $\overline{F}$ (where $F$ is $P'$, ONLY-IF($P''$), $Q'$ or $\neg Q''$). Now the partial completion $pcomp(P)$ of a program $P$ is $\overline{P'} \cup \overline{\text{ONLY-IF}(P'')} \cup CET$.

From Theorems 3.2 and 3.4 in (Stärk 1996) it follows that:

$$comp(P) \models_3 Q \quad \text{iff} \quad pcomp(P) \models \overline{Q'}$$
$$comp(P) \models_3 \neg Q \quad \text{iff} \quad pcomp(P) \models \overline{\neg Q''}$$

Let $F$ be $Q'$ (resp. $\neg Q''$). $pcomp(P) \models \overline{F}$ is equivalent to $P' \cup \text{ONLY-IF}(P'') \cup CET \models F$. □

### 4.3 Correctness of normal programs

We now introduce our method for proving program correctness. The presentation is followed by an example proof. Section 4.3.3 discusses completeness of the method and the next section compares the method with some other approaches. Section 4.3.5 shows how correctness w.r.t. approximate specifications can be employed in generalizing a known method of proving program termination. The reader may wish to skip Sections 4.3.3 – 4.3.5 in the first reading.

#### 4.3.1 Proof method

*Definition 4.5*
We say that a program $P$ is **correct** with respect to a specification $spec = (specS, specC)$ if for any query $Q$

(i) if $comp(P) \models_3 Q$ then $specS \cup specC' \models Q'$
(ii) if $comp(P) \models_3 \neg Q$ then $specS \cup specC' \models \neg Q''$

The reader may compare this definition with the informal discussion of Section 4.1, related to Definition 4.2.[11] In particular, if $P$ is correct with respect to $spec = (specS, specC)$, then from the soundness of SLDNF-resolution it follows that every computed answer $Q$ (of SLDNF-resolution) satisfies $specS \cup specC' \models Q'$. It means that for each positive literal $A$ in $Q$, $specS \models A$, and for each negative literal $\neg A$ in $Q$, $specC \models \neg A$. For $P$ and $spec$ as above, if a query $Q$ fails then $specS \cup specC' \models \neg Q''$ (by soundness of negation as failure). In the case of queries consisting of one literal $A$ (resp. $\neg A$) it means that $specC \models \neg A$ (resp. $specS \models A$).

---

[11] Notice that $specS \cup specC' \models Q'$ is equivalent to $I_S^4(spec) \models_4 Q$, and $specS \cup specC' \models \neg Q''$ is equivalent to $I_S^4(spec) \models_4 \neg Q$.

The same holds for any operational semantics, which is sound w.r.t. 3-valued completion semantics. This includes constructive negation (cf. (Drabent 1995) and the references therein) and extensions of SLDNF-resolution allowing selecting a non-ground negative literal $\neg A$ if $A$ fails or succeeds without binding its variables (Lloyd 1987; Stärk 1996).

The proposed proof method is given by the following theorem. ($spec_=$ is defined in Section 3.3.)

*Theorem 4.6* (*Correctness, normal programs*)
Let $P$ be a program and $spec = (specS, specC)$ a specification, such that

(a) $specS \cup specC' \models P'$
(b) $specS \cup specC' \cup spec_= \models \text{ONLY-IF}(P'')$

then

$P$ is correct w.r.t. $spec$.

*Proof*
From (a), (b) and $spec_= \models CET$ we obtain $specS \cup specC' \cup spec_= \models P' \cup \text{ONLY-IF}(P'') \cup CET$. Assume that $comp(P) \models_3 Q$ (respectively $comp(P) \models_3 \neg Q$). By Lemma 4.4, $specS \cup specC' \models Q'$ (resp. $specS \cup specC' \models \neg Q''$). □

### 4.3.2 Example correctness proof

We illustrate our correctness proof method by applying it to a program (from (Stärk 1996)) defining the subset relation. We present a detailed proof.

*Example 4.7*
Let $P$ be the following program:

$$subset(L, M) \leftarrow \neg notsubset(L, M)$$
$$notsubset(L, M) \leftarrow member(X, L), \neg member(X, M)$$
$$member(X, [X|L]) \leftarrow$$
$$member(X, [Y|L]) \leftarrow member(X, L)$$

Consider Herbrand specification $spec = (specS, specC)$, where

$$specS = sS_m \cup sS_n \cup sS_s, \qquad specC = sC_m \cup sC_n \cup sC_s$$

$$\begin{aligned}
sS_m &= \{member(x, l) \mid l \text{ is a list} \to x \in l\} \\
sC_m &= \{member(x, l) \mid l \text{ is a list} \land x \in l\} \\
sS_n &= \{notsubset(l, m) \mid l \text{ and } m \text{ are lists} \to l \nsubseteq m\} \\
sC_n &= \{notsubset(l, m) \mid l \text{ and } m \text{ are lists} \land l \nsubseteq m\} \\
sS_s &= \{subset(l, m) \mid l \text{ and } m \text{ are lists} \to l \subseteq m\} \\
sC_s &= \{subset(l, m) \mid l \text{ and } m \text{ are lists} \land l \subseteq m\}
\end{aligned}$$

($l \subseteq m$ means that all elements of $l$ are elements of $m$.)

We would like to prove that our program is correct with respect to the above specification *spec*. We show that conditions (a) and (b) of Theorem 4.6 are satisfied. This implies that whenever $subset(l,m)$ is a computed answer of $P$ then $sS_s \models subset(l,m)$, and thus if $l,m$ are lists then $l \subseteq m$. Also, whenever a query $subset(l,m)$ fails then $sC_s \models \neg subset(l,m)$. Hence $l$ or $m$ is not a list or $l \not\subseteq m$.

Let $spSC = specS \cup specC'$. In order to prove condition (a) one has to show, for each clause $C$ of $P$, that $spSC \models C'$. In order to prove condition (b) one has to show that each implication of ONLY-IF($P''$) is true in the interpretation $spSC \cup spec_=$.

Let us first consider the second clause of program $P$. For condition (a) we have to prove that:

$$spSC \models notsubset(L, M) \leftarrow member(X, L) \wedge \neg member'(X, M).$$

Let $l, m, x$ be any elements of the universe such that $spSC \models member(x,l) \wedge \neg member'(x,m)$. That means that $member(x,l) \in sS_m$ and $member(x,m) \notin sC_m$. We would like to prove that $notsubset(l,m) \in sS_n$. So assume that $l$ and $m$ are lists. From $member(x,l) \in sS_m$ we obtain that $x \in l$, and from $member(x,m) \notin sC_m$ — $x \notin m$. Hence $l \not\subseteq m$.

For condition (b) and predicate *notsubset* we have to show that

$$spSC \models notsubset'(L, M) \to \exists X (member'(X, L) \wedge \neg member(X, M))$$

Let $l, m$ be any elements of the universe such that $spSC \models notsubset'(l,m)$. So $l$ and $m$ are lists and $l \not\subseteq m$. So there exists an element, say $a$, such that $a \in l$ and $a \notin m$. Thus $member(a,l) \in sC_m$ and $member(a,m) \notin sS_m$. Hence $spSC \models member'(a,l) \wedge \neg member(a,m)$, so the implication above is true in $spSC$.

Let $C$ denote the first clause of $P$. Notice that $subset(L, M) \leftrightarrow \neg notsubset(L, M)$ is true both in $sS_s \cup sC_n$ and in $sC_s \cup sS_n$. After replacing *notsubset* by *notsubset'*, this implies $sS_s \cup sC_n' \models C'$, and hence (a) for the first clause. After replacing *subset* by *subset'*, this implies $sS_n \cup sC_s' \models subset'(L, M) \to \neg notsubset(L, M)$, hence (b) for predicate *subset*.

The proof for predicate *member* boils down to a proof of a definite program (a proof of correctness and part (i) of completeness proof, cf. Proposition 3.14). □

### 4.3.3 On completeness of the proof method

To discuss completeness of the proof method we need an ordering on specifications.

*Definition 4.8*
Let $sp = (spS, spC)$, $spec = (specS, specC)$ be specifications (over the same preinterpretation $\mathcal{J}$). We say that $sp$ is **stronger** than *spec* (written $sp \preceq spec$) if $spS \subseteq specS$ and $specC \subseteq spC$.

The set of atoms allowed to succeed (fail) by the stronger specification is a subset of the set of atoms allowed to succeed (fail) by the weaker one. The set of atoms required to succeed (fail) by the stronger specification is a superset of the analogical set for the weaker one.

Notice that this definition corresponds to an intuitive notion of a stronger specification. Let $spec_1 \preceq spec_2$, then for any program $P$, if $P$ is correct w.r.t. $spec_1$ then it is correct w.r.t. $spec_2$.

The ordering $\preceq$ on specifications corresponds to the information content ordering $\leq_k$ (Fitting 1991) on 4-valued specifications for correctness (cf. the last paragraphs of Section 4.1): $spec_1 \preceq spec_2$ iff $I_S^4(spec_1) \leq_k I_S^4(spec_2)$. It also holds that $spec_1 \preceq spec_2$ iff $I_C^4(spec_2) \leq_k I_C^4(spec_1)$.

We say that a proof method is **complete** for $P$ if the following condition holds: if $P$ is correct w.r.t. a specification $spec$ then there exists a specification stronger than $spec$ which satisfies the conditions of the proof method.

As the following examples show, the proof method for program correctness (Theorem 4.6) is not complete.

*Example 4.9*
Let $P$ be the following program:

$$p(f(x)) \leftarrow p(x)$$
$$q \leftarrow p(x)$$

Consider a non-proper Herbrand specification $spec = (\emptyset, \{q\})$, which says that no atom is allowed to succeed and all atoms except $q$ are allowed to fail. Notice that $I_S^4(spec) = \Phi_P \uparrow \omega$ and $\Phi_P \uparrow \omega$ is not the $\leq_k$-least fixpoint, where $\Phi_P$ is the 4-valued immediate consequence operator of $P$ (Fitting 1991). Program $P$ is correct w.r.t. this specification. Unfortunately neither $spec$ nor any specification stronger than $spec$ satisfies conditions (a) and (b) of Theorem 4.6 (for justification see below). Thus the proof method cannot be applied.

On the other hand program $P$ is correct w.r.t. a proper specification $(\emptyset, \emptyset)$, corresponding to the least fixpoint of the operator $\Phi_P$, and this specification does satisfy conditions (a) and (b). □

*Example 4.10*
Consider a program $P$:

$$p(a) \leftarrow$$
$$q \leftarrow \neg p(x)$$

and assume that the underlying language has exactly one function symbol $a$. (This example can be easily generalized for any finite set of function symbols.) Notice that $comp(P) \not\models_3 q$ and $comp(P) \not\models_3 \neg q$. Consider a Herbrand specification $spec = (\{p(a)\}, \{p(a), q\})$, which allows $q$ neither to succeed nor to fail. $P$ is correct w.r.t. $spec$ but the verification condition of our method (Theorem 4.6) is not satisfied. The latter is due to $\{p(a), p'(a), q'\} \not\models q' \rightarrow \exists x \neg p(x)$.

The condition holds for a weaker specification $(\{p(a)\}, \{p(a)\})$ that corresponds to the least fixpoint of $\Phi_P$ over one element Herbrand algebra.

To explain the incompleteness we will refer to the 4-valued immediate consequence operator $\Phi_P^{\mathcal{J}}$ (over a preinterpretation $\mathcal{J}$). A 4-valued interpretation $I$ is a *pre-fixpoint* of $\Phi_P^{\mathcal{J}}$ iff $\Phi_P^{\mathcal{J}}(I) \leq_k I$. For any $\alpha$, the interpretation $\Phi_P^{\mathcal{J}} \uparrow \alpha$ is 3-valued and is identical to the corresponding power of the 3-valued immediate consequence operator (Stärk 1996).

Conditions (a) and (b) of Theorem 4.6 mean that the 4-valued interpretation $I_S^4(spec)$ corresponding to a specification *spec* over a preinterpretation $\mathcal{J}$ is a prefixpoint of $\Phi_P^{\mathcal{J}}$. A program $P$ may be correct w.r.t. a specification *spec* for which $I_S^4(spec)$ is not preceded by the least fixpoint of $\Phi_P^{\mathcal{J}}$ (in the ordering $\leq_k$). Then there does not exist a specification which satisfies (a), (b) and is stronger than *spec*. Hence Theorem 4.6 is inapplicable in such a case. In Example 4.9 this happens because $\Phi_P^{\mathcal{J}}\uparrow\omega$ is not a fixpoint of $\Phi_P^{\mathcal{J}}$. In Example 4.10 the program is correct w.r.t. a specification, which is stronger than the $\Phi_P^{\mathcal{J}}\uparrow\omega$.

When the set of function symbols of the underlying language is infinite then the strongest Herbrand specification with respect to which $P$ is correct is $spec_\omega$, where $I_S^4(spec_\omega) = \Phi_P\uparrow\omega$. The latter follows from the fact (Kunen 1987) that $comp(P) \models_3 F$ iff $\Phi_P\uparrow n \models_3 F$ for some finite $n$ (where $F$ is a query or a negation of a query).

This reasoning can be summarized as:

*Proposition 4.11*
The correctness proof method of Theorem 4.6 is complete for an arbitrary program $P$ and for any specification weaker than the $\leq_k$-least fixpoint of $\Phi_P^{\mathcal{J}}$.

When the set of function symbols of the underlying language is infinite then the proof method is complete for an arbitrary Herbrand specification *spec* and any program $P$ for which $\Phi_P\uparrow\omega$ is the $\leq_k$-least fixpoint of $\Phi_P$.

We believe that cases for which the method is not complete are rather artificial (like those from the two examples above) and are rare in practice.

### *4.3.4 Correctness proving methods — comparison*

In this section we compare the correctness proof method from Section 4.3.1 with that for definite programs (Section 3.1) and with the approach of (Pedreschi and Ruggieri 1999). We show that the latter is (strictly) weaker, as far as declarative properties of programs are concerned.

We first show that the natural method for proving correctness of definite programs (Proposition 3.2) is a special case of the method for normal programs (Theorem 4.6). Let $P$ be a definite program and *specS* a specification for correctness. Take $spec = (specS, \emptyset)$. Then condition (a) is equivalent to $specS \models P$ (i.e. the verification condition of the natural method), and condition (b) reduces to $spec_= \models \text{ONLY-IF}(P)$, which trivially holds.

A straightforward way of generalizing to normal programs the natural method for proving correctness of definite programs is to replace 2-valued interpretations by 4-valued ones, and programs by program completions:

*Proposition 4.12*
Let $P$ be a program, $Q$ a query and $I$ a 4-valued interpretation such that $I \models_4 comp(P)$.

1. If $comp(P) \models_3 Q$ then $I \models_4 Q$. If $comp(P) \models_3 \neg Q$ then $I \models_4 \neg Q$.
2. $P$ is correct w.r.t. a specification *spec* such that $I_S^4(spec) = I$.

*Proof*
Let $F$ be $Q$ or $\neg Q$. From $comp(P) \models_3 F$ it follows that $comp(P) \models_4 F$ (Stärk 1996). As $I \models_4 comp(P)$, we obtain $I \models_4 F$. If $I_S^4(spec) = I$ then implications 1. are equivalent to the conditions (i), (ii) of Definition 4.5 of program correctness. □

The proof method provided by Proposition 4.12 is in fact weaker than that of Theorem 4.6, as $I_S^4(spec) \models_4 comp(P)$ implies conditions (a) and (b) of the Theorem but not vice versa. This is because $I_S^4(spec) \models_4 comp(P)$ means that $I_S^4(spec)$ is a fixpoint of the 4-valued immediate consequence operator $\Phi_P$ (Stärk 1996), while conditions (a), (b) mean that it is a pre-fixpoint of $\Phi_P$. (For details see Section 4.3.3.)

Pedreschi and Ruggieri (1999) presented an operational method for proving total correctness of normal programs. It is an extension of the method for definite programs discussed in Section 3.2. The method uses call-success specifications (cf. Section 3.2), the difference is that the pre- and postconditions are Herbrand interpretations. The core of the method is the following definition of a proof relation $\vdash_t$. A **level mapping** is a function from ground atoms to natural numbers. For a level mapping $|\ |$ and an atom $A$, $|A|$ will denote the maximum of $\{|A\theta| : A\theta \text{ is ground}\}$ or $\infty$ when such maximum does not exist.

*Definition 4.13* (Pedreschi and Ruggieri 1999, Definition 5.3)
Let $P$ be a program, and $\langle Pre, Post \rangle$ a call-success specification, where $Pre, Post$ are Herbrand interpretations. We write $\vdash_t \{Pre\}P\{Post\}$ iff there exists a level mapping $|\ |$ such that:
(i) for every ground instance $A \leftarrow L_1, \ldots, L_n$ of a clause of $P$:
1. for $i \in [1, n]$:
$$Pre \models A \ \wedge \ Post \models L_1, \ldots L_{i-1} \implies$$
$$\begin{cases} Pre \models B_i \ \wedge \ |A| > |B_i| & \text{if } L_i = B_i \\ Pre \models B_i \ \wedge \ |A| > |B_i| & \text{if } L_i = \neg B_i \end{cases}$$
2. $Pre \models A \ \wedge \ Post \models L_1, \ldots, L_n \implies Post \models A$
(ii) $T_P(Post) \supseteq Post \cap Pre$.

If $\vdash_t \{Pre\}P\{Post\}$ holds and $P$ does not flounder for (ground atomic) queries from $Pre$ then $P$ is totally correct w.r.t. $\langle Pre, Post \cap Pre \rangle$ in the following sense:

1. $Post \cap Pre$ is the set of those atoms from $Pre$ that succeed,
2. $Pre \setminus Post$ is the set of those atoms from $Pre$ that fail,
3. if $Pre \models A_0$, $|A_0|$ is finite, and $L$ is $A_0$ or $\neg A_0$ then the LDNF-tree with the root $L$

   (a) is finite, and
   (b) for each selected literal $A$ or $\neg A$ in the tree, $Pre \models A$.

Properties 1. and 2. mean that $P$ is correct (in the sense of Definition 4.5) w.r.t. a specification *spec* for which $(\mathcal{H} \setminus Pre) \cup (Post \cap Pre)$ are the ground atoms allowed to succeed and $(\mathcal{H} \setminus Pre) \cup (Pre \setminus Post)$ are the ground atoms allowed to fail. Simple calculation results in $spec = ((\mathcal{H} \setminus Pre) \cup Post,\ Pre \cap Post)$.

The following proposition states that whenever the method of (Pedreschi and Ruggieri 1999) can provide a proof of program correctness (in the sense of Definition 4.5) then a proof can be obtained by our method.

*Proposition 4.14*
Assume that $\vdash_t \{Pre\}P\{Post\}$ holds. Then conditions (a) and (b) of Theorem 4.6 hold for specification $spec = (specS, specC)$, such that $specS = (\mathcal{H} \setminus Pre) \cup Post$ and $specC = Pre \cap Post$. Hence $P$ is correct w.r.t. *spec*.

*Proof*
To prove (a) consider a ground instance $H \leftarrow \vec{B}$ of a clause of $P$ and assume that $specS \cup specC' \models \vec{B}'$ and $Pre \models H$. (If $Pre \not\models H$ then condition (a) trivially holds.) Then for each literal $L$ of $\vec{B}$ we obtain from (i) 1. by induction that $Pre \models L$ if $L$ is positive, $Pre \models \neg L$ if $L$ is negative, and hence $Post \models L$ (as in the first case $specS \models L$ and in the second case $specC \models L$). Thus $Post \models H$, by (i) 2., and $specS \cup specC' \models H$.

To prove (b) it is sufficient to show that for every ground atom $A$ such that $specC \models A$ there exists a ground instance of a program clause $A \leftarrow \vec{B}$ such that $specS \cup specC' \models \vec{B}''$. Let $A$ be a ground atom for which $specC \models A$ (i.e. $A \in Pre \cap Post$). By (ii) $A \in T_P(Post)$. From the definition of $T_P$ it follows that there exists a ground instance of a program clause $A \leftarrow \vec{B}$ such that $Post \models \vec{B}$. From (i) 1. each literal $L_i$ of $\vec{B}$, where $L_i = B_i$ or $L_i = \neg B_i$, satisfies $Pre \models B_i$. In case $L_i = B_i$ we have $Post \models L_i$ and $Pre \models L_i$, thus $specC \models L_i$. In case $L_i = \neg B_i$ we have $Post \models L_i$ and $Pre \not\models L_i$, thus $specS \models L_i$. Hence $specS \cup specC' \models L_i''$ for every literal $L_i$ of $\vec{B}$, so $specS \cup specC' \models \vec{B}''$. □

The condition $\vdash_t \{Pre\}P\{Post\}$ in the Proposition may be weakened: notice that no facts concerning the level mapping (from the definition of $\vdash_t$) were used in the proof.

We showed that the proof method of (Pedreschi and Ruggieri 1999) is weaker than that of Theorem 4.6, as far as the declarative semantics and program correctness are concerned. It is actually strictly weaker, due to the following limitations. The method deals only with total correctness, thus correctness, completeness and termination have to be proved together; none of them can be dealt with separately. As a result, the method is not applicable to approximate specifications, for which a program is correct but not complete (or does not terminate). A specification has to be exact for the atoms in $Pre$, in the sense that it states exactly which of them succeed and which fail. Formally, $Pre \cap Post$ is unique for a given program and precondition.[12] The method deals with LDNF-resolution and is inapplicable to programs that flounder or do not terminate under Prolog selection rule (but are intended to be executed under some other operational semantics, for instance delays or constructive negation). It is also inapplicable when the LDNF-tree is infinite but has success nodes (e.g. when a query has an infinite set of computed answers). In all these cases the method of Theorem 4.6 is applicable. Notice also that the operational method, in contrast to our approach, requires separate proving of non-floundering.

---

[12] Notice that this restriction does not concern the operational method for definite programs (Section 3.2); that method permits approximate specifications.

Obviously, the method of (Pedreschi and Ruggieri 1999) deals also with properties which are out of the scope of our approach, namely termination and the form of selected literals in LDNF-resolution. See Section 4.4 for a comparison with the method for proving program completeness proposed in this paper.

### 4.3.5 A note on proving termination

This section, in contrast to the rest of the paper, considers a property which is not declarative. We show how approximate specifications (which describe declarative notion of program correctness) can be employed in generalizing a well-known method of proving termination.

Apt and Pedreschi (1993) presented a method of proving termination of normal programs with the Prolog selection rule. They introduced a notion of an *acceptable program* (w.r.t. a 2-valued interpretation $I$ and a level mapping). Any acceptable program is left terminating, this means it terminates with the Prolog selection rule for all ground goals. The interpretation $I$ is a model of the program and a model of $comp(P^-)$, where $P^-$ is the involved in negation part of the program.[13] It turns out that $I$ is unique for all the predicates in $P^-$ (Apt and Pedreschi 1993). This is a disadvantage of the approach; to show that $P$ is left terminating one has to know the unique interpretation.

We show that the unique 2-valued interpretation can be replaced by an approximate specification *spec* w.r.t. which the program is correct. We introduce a notion of an *approximately acceptable* program (*a-acceptable* in short). We prove that each a-acceptable program is acceptable, and thus left terminating.

*Definition 4.15*
Let $P$ be a program, $|\ |$ a level mapping, and $spec = (specS, specC)$ a specification. $P$ is called **a-acceptable** with respect to $|\ |$ and *spec* if $P$ is correct w.r.t. *spec*, and for every ground instance $A \leftarrow L_1, \ldots, L_n$ of a clause from $P$ the following implication holds for $i \in [1, n]$:

$$\text{if} \quad specS \cup specC' \models \bigwedge_{j=1}^{i-1} L'_j \quad \text{then} \quad |\,A\,| > |\,L_i\,|\,.$$

where $|\neg B| = |B|$ for any ground atom $B$.

---
[13] To define $P^-$, first $Neg_P^*$ is introduced; it is the least set such that any predicate symbol occurring in a negative literal of $P$ is in $Neg_P^*$, and if $p \in Neg_P^*$ occurs in the head of a clause $C$ of $P$ and $q$ in the body of $C$ then $q \in Neg_P^*$. Program $P^-$ contains those clauses of $P$ that use symbols from $Neg_P^*$ in their heads.

This definition differs from that of an acceptable program in two aspects. Condition $I \models \bigwedge_{j=1}^{i-1} L_j$ has been replaced by $specS \cup specC' \models \bigwedge_{j=1}^{i-1} L'_j$. Also, $P$ is required to be correct w.r.t. *spec*, instead of $I$ being a model of $P \cup comp(P^-)$, where *comp* is taken w.r.t. the alphabet $Neg_P^*$ of predicate symbols.[14]

The notion of a-acceptability is a generalization of that of acceptability. If $P$ is acceptable w.r.t. $|\ |$ and an interpretation $I$, which is a model of $comp(P)$, then $P$ is a-acceptable w.r.t. $|\ |$ and $spec = (I, I)$. This follows directly from the definitions and from the fact that $I \models comp(P)$ implies correctness of $P$ w.r.t. $(I, I)$.

In a general case, if $P$ is acceptable w.r.t. $|\ |$ and an interpretation $I$ over a preinterpretation $\mathcal{J}$ then $P$ is a-acceptable w.r.t. $|\ |$ and $spec = (I, I \setminus \mathcal{B})$, where $\mathcal{B}$ is the set of $\mathcal{J}$-atoms of the form $p(\vec{t})$, where $p \notin Neg_P^*$. This follows from the definitions and from the fact that $I \models P \cup comp(P^-)$ implies that $P$ is correct w.r.t. $(I, I \setminus \mathcal{B})$. The latter implication follows from a lemma that for any $\mathcal{J}$-literal $L$ and any natural number $n$, if $L$ is true in (the 3-valued interpretation) $\Phi_P^{\mathcal{J}} \uparrow n$ then $L'$ is true in (2-valued) $I \cup (I \setminus \mathcal{B})'$, where $\Phi_P^{\mathcal{J}}$ is the 3-valued immediate consequence operator over $\mathcal{J}$. We skip details of the proof.

To show that a-acceptability implies left termination we employ the following theorem, analogous to Theorem 6.7 of (Apt and Pedreschi 1993). In what follows, $\Phi_P$ denotes the 3-valued immediate consequence operator. A 3-valued Herbrand interpretation is total if any ground atom is either true or false in this interpretation.

*Theorem 4.16*
Assume that the set of function symbols is infinite. Let $P$ be an a-acceptable program w.r.t. $|\ |$ and *spec*. Then $\Phi_P \uparrow \omega$ is total.

*Proof (outline)*
The proof is basically the same as that of (Apt and Pedreschi 1993), however a substantial part of the latter proof is made shorter. The part is entitled Subcase 2 and shows that $\Phi_P \uparrow n \models_3 \neg L_{\bar{k}}$, for a literal $L_{\bar{k}}$ which is not undefined in $\Phi_P \uparrow n$ and for which $I \models \neg L_{\bar{k}}$ (where $P$ is acceptable w.r.t. $I$). In our case $I \models \neg L_{\bar{k}}$ is replaced by $specS \cup specC' \models \neg L'_{\bar{k}}$ (where $spec = (specS, specC)$) and the whole Subcase 2 is reduced to the following: an assumption that $L_{\bar{k}}$ is true in $\Phi_P \uparrow n$ leads (by Kunen theorem (Kunen 1987; Doets 1994)) to $comp(P) \models_3 L_{\bar{k}}$ and, by correctness of $P$, to $specS \cup specC' \models L'_{\bar{k}}$, contradiction. So $\Phi_P \uparrow n \models_3 \neg L_{\bar{k}}$. □

Now we can prove the main result.

*Theorem 4.17*
Assume that the set of function symbols is infinite. Let $P$ be an a-acceptable program w.r.t. $|\ |$ and *spec*. Then $P$ is acceptable (w.r.t. $|\ |$ and some interpretation) and $P$ is left terminating.

---

[14] Thus if $comp(P^-)$ contains an axiom $\neg p(\vec{x})$ then $p \in Neg_P^*$.

*Proof*
Let $spec = (specS, specC)$ and $P$ be as in the assumptions of the theorem. By Theorem 4.16 $\Phi_P\uparrow\omega$ is total, thus this is the least fixpoint of $\Phi_P$, hence $\Phi_P\uparrow\omega \models_3 comp(P)$. Let $I$ be the ground atoms true in $\Phi_P\uparrow\omega$, we have $I \models comp(P)$.

Now to show that $P$ is acceptable w.r.t. $I$ and $|\ |$, it is sufficient to show that $I \models L$ implies $specS \cup specC' \models L'$, for any ground literal $L$.

Consider a ground literal $L$ such that $I \models L$. This means $\Phi_P\uparrow\omega \models_3 L$. Thus $\Phi_P\uparrow n \models_3 L$ for some $n < \omega$. Hence $comp(P) \models_3 L$, by Kunen theorem (Kunen 1987; Doets 1994). So $specS \cup specC' \models L'$, by the correctness of $P$. □

*Example 4.18*
To illustrate the method we employ the following program, similar to program GAME of (Apt and Pedreschi 1993).

$$win(X) \leftarrow move(X,Y), \neg win(Y).$$
$$move([l|X], X).$$
$$move([l,l|X], X).$$
$$move([l,l,l,l|X], X).$$

It models a two person game, in each move a player removes a certain number of tokens, the one who removes the last token wins.

Let $|\ |$ be the function on ground terms such that $|f(t_1,\ldots,t_n)| = 0$ if $f \neq [\ |\ ]$ and $|[t_1|t]| = |t|+1$. Consider a level mapping $|win(t)| = |t|+1$, $|move(t_1,t_2)| = |t_1|$, and a Herbrand specification $spec = (sS_w \cup sS_m, \emptyset)$ where

$$sS_w = \{\ win(t) \mid t \text{ is a ground term }\},$$
$$sS_m = \{\ move(t_1, t_2) \mid |t_1| > |t_2|\ \}.$$

The program is obviously correct w.r.t. *spec*, and it is easy to check that it is a-acceptable. Thus it is left terminating. Notice that the analogical proof in (Apt and Pedreschi 1993) requires providing the unique model $I$ of the program completion. This model is also needed to apply the method of (Pedreschi and Ruggieri 1999) for this program. (More precisely, in order to show termination for goals from a precondition *Pre*, one has to know $I \cap Pre$; see the discussion of Section 4.3.4.)

The proof of (Apt and Pedreschi 1993) considers arbitrary relation *move* for which the corresponding graph is finite and acyclic. Our proof can be easily adjusted to such case, by replacing function $|\ |$ above by the function $f$ used in (Apt and Pedreschi 1993) to define the level mapping in the proof.[15]

Left termination does not imply a-acceptability. But according to Theorem 4.18 of (Apt and Pedreschi 1993) if $P$ is a left terminating, non-floundering program then $P$ is acceptable w.r.t. some level mapping $|\ |$ and a model $I$ of $comp(P)$. Thus $P$ is a-acceptable w.r.t. $|\ |$ and specification $spec = (I, I)$.

In this section we generalized the notion of acceptable program, so that in termination proofs one can use approximate specifications instead of unique models. We deal with termination of a program for all ground goals. There is another path of generalizing the method of (Apt and Pedreschi 1993) to programs that terminate only for some ground goals (Bossi et al. 1994; Schreye et al. 1992). The approach of (Pedreschi and Ruggieri 1999) belongs to this path. It proves left termination for the goals satisfying the precondition, however — as discussed in Section 4.3.4 — for such goals the specification has to be exact.

---

[15] As function $f$ is defined only for the nodes of the graph, it should be generalized to arbitrary terms. This can be done by assuming $f(t) = 0$ for any term which is not a node of the graph.

We expect that some improvements of the method of (Apt and Pedreschi 1993) are also applicable to the method presented here, for instance a weakening of inequalities in case of non-mutually recursive predicates (Apt and Pedreschi 1994; Pedreschi and Ruggieri 1999).

### *4.4 Completeness of normal programs*

As the operational semantics for normal programs we assume SLDNF-resolution, as defined by Apt and Doets (1994). To discuss completeness we need to refer to the notion of SLDNF-tree. We outline its definition below, for more details the reader is referred to (Apt and Doets 1994) or (Doets 1994).

An SLDNF-tree for query $Q$ and program $P$ is a set of trees, with one of them distinguished as the main tree. The nodes of the trees are queries and the trees are, roughly speaking, SLDNF-trees of (Lloyd 1987). $Q$ is the root of the main tree. Any node with a non-ground negative literal selected is a leaf of a tree, such a node is marked *floundered*. Whenever a ground negative literal $\neg A$ is selected in a node $N$ then there exists a subsidiary tree with the root $A$. The whole SLDNF-tree may be viewed as a tree of trees, in which the tree with the node $N$ is the parent of the subsidiary tree with the root $A$.

The leaves of each tree can be marked *failed* or *success*, with the expected meaning. So if a leaf $N$ is neither marked *failed* nor *success* then a negative literal $\neg A$ is selected in $N$, moreover $A$ is non-ground or the subsidiary tree for $A$ neither succeeds nor finitely fails. A tree succeeds if it has a *success* leaf. A tree finitely fails if it is finite and all its leaves are marked *failed*.

The SLDNF-tree succeeds (finitely fails) if the main tree does. To each success leaf of the main tree there corresponds a computed answer substitution $\theta$ for $Q$ (and a computed answer $Q\theta$), defined as expected.

#### *4.4.1 Proof method*

In this section we introduce a method for proving program completeness. Then we briefly discuss completeness of the method and provide a comparison with an operational proof method.

*Definition 4.19*
We say that a program $P$ is **complete for a query** $Q$ w.r.t. a specification $spec = (specS, specC)$ if

(i) $specS \cup specC' \models Q''$ implies $comp(P) \models_3 Q$,
(ii) $specS \cup specC' \models \neg Q'$ implies $comp(P) \models_3 \neg Q$.

Program $P$ is **complete** w.r.t. $spec$ if it is complete for any query $Q$.[16]

We say that a program $P$ is **SLDNF-complete** for a query $Q$ w.r.t. a specification $spec = (specS, specC)$ if

(i) $specS \cup specC' \models Q''\sigma$ implies that some SLDNF-tree for $Q$ succeeds with an answer $Q\theta$ more general than $Q\sigma$,
(ii) $specS \cup specC' \models \neg Q'$ implies that some SLDNF-tree for $Q$ finitely fails.

From the soundness of SLDNF-resolution it follows that SLDNF-completeness implies completeness.

For Herbrand specifications completeness implies correctness:

*Proposition 4.20*
If a program $P$ is complete w.r.t. a Herbrand specification $spec = (specS, specC)$ then

1. $spec$ is proper, and
2. $P$ is correct w.r.t. $spec$.

*Proof*
1. If $spec$ is not proper then there exists a ground atom $A \in specC \setminus specS$. By Definition 4.19, $comp(P) \models_3 A$ and $comp(P) \models_3 \neg A$; contradiction.
2. Assume $P$ is not correct w.r.t. $spec$. So for some query $Q$ we have $comp(P) \models_3 Q$ and $specS \cup specC' \not\models Q'$, or $comp(P) \models_3 \neg Q$ and $specS \cup specC' \not\models \neg Q''$. In the first case we have $specS \cup specC' \models \neg Q'\theta$ for some ground $Q\theta$ and, by Definition 4.19, $comp(P) \models_3 \neg Q\theta$; contradiction. Similarly in the second case, $specS \cup specC' \models Q''\theta$ and $comp(P) \models_3 Q\theta$; contradiction. □

To show that the proposition does not hold for non-Herbrand specifications, consider a preinterpretation $\mathcal{J}$, the set $S = \{\, t \in |\mathcal{J}| \mid t \text{ is a value in } \mathcal{J} \text{ of a ground term}\,\}$, and an element $u \in |\mathcal{J}| \setminus S$. Let $I = \{\, p(t) \mid t \in S \,\}$. Program $\{\, p(X) \leftarrow \,\}$ is complete w.r.t. $(I, I)$ and w.r.t. $(I, I \cup \{p(u)\})$, the latter specification is not proper. The program is however not correct w.r.t. any of these specifications. We consider completeness w.r.t. non-proper specifications as a rather pathological case.

The following theorem gives sufficient conditions for program completeness.

*Theorem 4.21 (Completeness, normal programs)*
Assume that the set of function symbols is infinite. Let $P$ be a program, $Q$ a query and $spec = (specS, specC)$ a specification such that

1. $P$ is correct w.r.t. $spec$, and
2. there exists an SLDNF-tree for $Q$ such that its main tree is finite and all the leaves of the main tree are marked *failed* or *success*.

Then $P$ is SLDNF-complete and complete for $Q$ w.r.t. $spec = (specS, specC)$.

---

[16] These notions can be expressed in terms of the 4-valued logic by using the fact that $spec \cup specC' \models Q''$ is equivalent to $I_C^4(spec) \models_4 Q$, and $specS \cup specC' \models \neg Q'$ is equivalent to $I_C^4(spec) \models_4 \neg Q$.

Condition 2. implies that each $\neg A$ selected in the main tree is ground and the subsidiary tree for $A$ succeeds or fails. Notice that the SLDNF-tree may be infinite or contain floundering nodes. However the "important part" of it is finite and without floundering and can be computed under some search strategy in a finite number of steps. (When a success is obtained in a subsidiary tree, traversing this tree can be abandoned.) Remember that the selection rule in the tree is arbitrary, this includes delay mechanisms. If due to delays no literal is selected in a node of the main tree then condition 2. is not satisfied. It is a kind of floundering, the node is a leaf marked neither *failed* nor *success*.[17]

The proof uses the following lemma.

*Lemma 4.22*
Assume that the set of function symbols is infinite. Let $t, t_1, \ldots, t_n$ be terms such that $t$ is not an instance of any $t_i$. Then there exists an instance of $t$ which is not unifiable with any $t_i$.

*Proof*
Let $v_1, \ldots, v_k$ ($k \geq 0$) be the variables of $t$. Let $f_1, \ldots, f_k$ be distinct function symbols (of arity $\geq 0$) not occurring in $t, t_1, \ldots, t_n$ and $c$ be a constant. Let $u_i = f_i(c, \ldots, c)$, for $i = 1, \ldots, k$, be terms. Consider a substitution $\theta = \{v_1/u_1, \ldots, v_k/u_k\}$. Assume that $s = t\theta$ is unifiable with some $t_i$. As $s$ is ground, $t\theta = t_i\sigma$ for some substitution $\sigma$. Terms $u_1, \ldots, u_k$ occur (as subterms) in $\sigma$ (since they occur in $t_i\sigma$). Let us replace each occurrence of term $u_j$ in $\sigma$ by the variable $v_j$, obtaining $\sigma'$, and remove all the pairs of the form $v_j/v_j$ from $\sigma'$, obtaining a substitution $\sigma''$. Then $t = t_i\sigma''$, contradiction with the assumption of the lemma. So $s = t\theta$ is not unifiable with any $t_i$.  $\square$

Now we can present a proof of Theorem 4.21.

*Proof*
Assume that conditions 1. and 2. hold. Let **T** be the SLDNF-tree for $Q$ satisfying 2.

**(i)** Let $specS \cup specC' \models \neg Q'$. We show that **T** finitely fails. Assume that it does not. Then **T** succeeds with some answer $Q\theta$. By correctness we have $specS \cup specC' \models Q'\theta$. Contradiction (with $specS \cup specC' \models \neg Q'$).

**(ii)** Let $specS \cup specC' \models Q''\sigma$. We want to show that **T** contains an answer more general than $Q\sigma$. Assume the contrary. Let $Q_1, \ldots, Q_n$ ($n \geq 0$) be the answers of **T**, $Q\sigma$ is not an instance of any of them. By Lemma 4.22, there exists $Q\sigma\theta$ which is not unifiable with any of $Q_1, \ldots, Q_n$.

Consider the SLDNF-tree **U** for $Q\sigma\theta$ under the same selection rule[18] as **T**. The nodes of **U** are instances of corresponding nodes of **T**. The main tree of **U** is finite and all its leaves are marked *failed* or *success*. Any answer of **U** is an instance of some $Q_i$ and of $Q\sigma\theta$. So no such answer exists and **U** finitely fails. Hence $specS \cup specC' \models \neg Q''\sigma\theta$. Contradiction.  $\square$

---

[17] Formally, the tree is not an SLDNF-tree, as a selection rule has to select a literal in every non-empty query.
[18] We omit tedious formalization of this notion.

We conclude this section by discussing completeness of the method and comparing it with the method of (Pedreschi and Ruggieri 1999).

A program $P$ may be complete w.r.t. a specification without satisfying condition 2. of Theorem 4.21 (e.g. a query may succeed but the main tree may be infinite). In such cases the method of proving completeness is inapplicable. However the method is trivially complete for Herbrand specifications and programs which terminate without floundering for the considered queries (formally: satisfy condition 2. of the theorem). This is because completeness w.r.t. a Herbrand specification implies correctness w.r.t. the same specification (Proposition 4.20). Hence condition 1. of Theorem 4.21 holds.

The operational proof method for total correctness of Pedreschi and Ruggieri (1999) was discussed in Section 4.3.4. Total correctness includes completeness. From the discussion after Definition 4.13 it follows immediately that the verification condition of that method implies completeness of $P$ (in the sense of Definition 4.19) for ground atomic queries from $Pre$ with respect to specification $spec = ((\mathcal{H} \setminus Pre) \cup Post,\ Pre \cap Post)$. It also implies termination for such queries and, by Proposition 4.14, correctness of $P$ w.r.t. $spec$. Thus the completeness can be shown by our method, as the conditions of Theorem 4.21 hold (for all ground atomic queries which are requested by $spec$ to succeed or fail). Remember however that our approach requires a separate termination proof.

This shows that for ground atomic queries the method of this section is stronger than that of (Pedreschi and Ruggieri 1999) (as far as program completeness in the sense of Definition 4.19 is concerned). It is also strictly stronger, as it applies to programs that loop or flounder under Prolog selection rule, but do not under some other one.[19]

### 4.4.2 Example completeness proof

Let us illustrate our method of proving completeness of normal programs by applying it to a program defining the subset relation with an additional requirement that a subset must be a list without repetitions. The example is rather lengthy, as our intention was to present a detailed proof.

*Example 4.23*
Let $P$ be the following program:

$$subs([\,], L) \leftarrow$$
$$subs([H|T], LH) \leftarrow select(H, LH, L), subs(T, L), \neg member(H, T)$$
$$select(H, [H|L], L) \leftarrow$$
$$select(H, [X|L], [X|LH]) \leftarrow select(H, L, LH)$$

The definition and specification of *member* are the same as in Example 4.7. A

---

[19] A wider class of queries is dealt with by Theorem 5.10 of (Pedreschi and Ruggieri 1999). It provides a criterion implying condition (i) of Definition 4.19 of completeness. Also in this case the criterion implies that a completeness proof by our method exists.

Herbrand specification for $P$ is $spec = (specS, \ specC)$, where

$$specS = sS_m \cup sS_{sel} \cup sS_{subs}, \quad specC = sC_m \cup sC_{sel} \cup sC_{subs}$$

$$\begin{aligned}
sS_{sel} &= \{select(e,l,m) \mid l \text{ is a list} \rightarrow e \in l \wedge m \text{ is a list} \wedge l \approx [e|m]\} \\
sC_{sel} &= \{select(e,l,m) \mid l \text{ and } m \text{ are lists such that} \\
&\quad l = [e_1, \ldots, e_i, e, e_{i+1}, \ldots e_k], \ m = [e_1, \ldots, e_i, e_{i+1}, \ldots e_k], 0 \leq i \leq k\} \\
sS_{subs} &= \{subs(l,m) \mid m \text{ is a list} \rightarrow listd(l) \wedge l \subseteq m\} \\
sC_{subs} &= \{subs(l,m) \mid m \text{ is a list} \wedge listd(l) \wedge l \subseteq m\}
\end{aligned}$$

Here $l \approx m$ means that lists $l$ and $m$ contain the same elements and $listd(l)$ means that $l$ is a list with distinct elements.

Let $spSC = specS \cup specC'$. To prove condition (a) for the second clause of predicate $subs/2$, assume that

$$spSC \models select(h,lh,l) \wedge subs(t,l) \wedge \neg member'(h,t). \qquad (A)$$

We show that $subs([h|t],lh) \in sS_{subs}$. So let $lh$ be a list. From $(A)$ it follows that:

(1) $select(h,lh,l) \in sS_{sel}$ hence $h \in lh$ and $l$ is a list such that $lh \approx [h|l]$;
(2) $subs(t,l) \in sS_{subs}$ hence $listd(t)$ and $t \subseteq l$, thus $[h|t] \subseteq lh$, by (1);
(3) $member(h,t) \notin sC_m$ hence $h \notin t$ (since $t$ is a list), thus $listd([h|t])$, by (2).

We obtain $[h|t] \subseteq lh$ and $listd([h|t])$, this completes the proof of condition (a) for the most complex clause of $P$.

Let us now prove condition (b) for predicate $subs/2$. We show that

$spSC \cup spec_= \models subs'(S,M) \rightarrow$
$S = [\,] \vee \exists H,T,L\,(S = [H|T] \wedge select'(H,M,L) \wedge subs'(T,L) \wedge \neg member(H,T)).$

Let $s, m$ be elements of the universe such that $spSC \models subs'(s,m)$, i.e. $subs(s,m) \in sC_{subs}$. So $m$ is a list, $s$ is a list of distinct elements and $s \subseteq m$. The case of $s = [\,]$ is obvious. Otherwise $s = [h|t]$. Since $h \in s$ and $s \subseteq m$, $m = [m_1, \ldots, m_{i-1}, h, m_{i+1}, \ldots, m_k]$. Take $[m_1, \ldots, m_{i-1}, m_{i+1}, \ldots, m_k]$ as $l$. Obviously $select(h,m,l) \in sC_{sel}$. From $listd([h|t])$ we have $listd(t)$ and $h \notin t$. Thus $member(h,t) \notin sS_m$. Also $subs(t,l) \in sC_{subs}$, as $[h|t] \subseteq m$. Thus the right hand side of the implication holds.

The remaining part of the proof of conditions (a), (b) is easier and is skipped here. It follows that $P$ is correct w.r.t. $spec$.

Consider a query $Q = subs(L,M)$, where $L$ is a variable and $M$ a ground list. Once it is shown that for such queries $P$ terminates without floundering (under some selection rule and search strategy), it follows that $P$ is complete for such queries. This means that for a given set all its subsets will be computed (i.e. all the permutations of the corresponding lists).

Assume that we do not have a termination proof and request all answers to a query $Q$ from an interpreter with run-time checks for floundering. Then if the execution terminates, we know that all the answers for $Q$ required by the specification have been produced. This happens in the case of our example program $P$ and Prolog. □

### *4.5 Example*

In this section we illustrate our method of proving correctness and completeness of normal programs by a larger example. We have chosen a program calculating the transitive closure of a given relation and its toy application to searching airway connections satisfying given requirements. Transitive closure is used as an example in many papers on proving program properties, e.g. (Apt 1995; Ferrand and Deransart 1993; Malfon 1994).

For our example we choose rather arbitrary approximate specifications. The purpose is to illustrate how just some of a program's properties can be proven. We also show how the method applies in case of extending a program by adding new predicates.

Information about the flights is given by predicate *direct*/3; *direct(from, to, flight(time, price))* denotes that there exists a direct flight from *from* to *to*, time of the flight is equal to *time* and its cost is *price*. We assume that *time* and *price* are natural numbers.

Let FLIGHTS be the following program:

$good\_conn(X, Y, Req) \leftarrow connect(X, Y, Dxy), satisfies(Dxy, Req)$

$connect(X, Y, Dxy) \leftarrow connect(X, Y, Dxy, [X])$

$connect(X, Y, [D], V) \leftarrow direct(X, Y, D)$
$connect(X, Z, [D|Dyz], V) \leftarrow$
    $direct(X, Y, D),$
    $\neg member(Y, V),$
    $connect(Y, Z, Dyz, [Y|V])$

% *satisfies(ListOfFlightsInfo, Requirements), where*
%   *Requirements = req(MaxNoOfTransfers, MaxTotalCost, MaxFlightTime)*

$satisfies(List, req(MaxTr, MaxTotalCost, MaxFlightTime)) \leftarrow$
    $analyze(List, NoOfTransfers, TotalCost, MaxFlightTime),$
    $lesseq(NoOfTransfers, MaxTr),$
    $lesseq(TotalCost, MaxTotalCost)$

% *analyze(ListOfFlightsInfo, NumberOfTransfers, Cost, MaxFlightTime)*

$analyze([\,], -1, 0, MaxFlightTime) \leftarrow$
$analyze([flight(Time, Price)|List], NoOfTransfers, Cost, MaxFlightTime) \leftarrow$
    $lesseq(Time, MaxFlightTime),$
    $analyze(List, NoOfTrL, ListCost, MaxFlightTime),$
    $add(ListCost, Price, Cost),$
    $add(NoOfTrL, 1, NoOfTransfers)$

We omit here the definition of *direct*/3 and definitions of (built-in) predicates *lesseq*/2 and *add*/3. The definition of *member*/2 is the same as in the previous examples.

Predicate *direct*/3 defines a directed graph: *direct(x, y, info)* means that there is an edge from *x* to *y* labelled by *info*. We assume that there is no loop, i.e. an edge (direct flight) from $x$ to $x$. Following (Ross and Wright 2003) by a path we will understand a non-empty sequence of edges satisfying standard condition: the terminal vertex of each edge is the initial vertex of the next. A cycle is a path $(x_1, x_2), \ldots, (x_n, x_1)$, $n \geq 1$, such that $x_1, \ldots, x_n$ are distinct vertices. A graph that contains no cycle is called acyclic. A path is acyclic if the graph consisting of the vertices and edges of the path is acyclic.

We begin with proving correctness and completeness of FLIGHTS with respect to the specification such that its first part ($specS$) is constructed from the following sets of ground atoms (which may be called predicate specifications). Let $G$ denote the labelled graph defined by $direct/3$, and $\mathcal{N}, \mathcal{Z}$ the sets of natural and integer numbers, respectively.

$$
\begin{aligned}
specS_{good\_conn} &= \{\, good\_conn(x, y, req(k, c, t)) \mid \text{there exists in } G \text{ a path} \\
&\qquad \text{from } x \text{ to } y \text{ such that the total cost of the connection} \\
&\qquad \text{does not exceed } c, \quad c \in \mathcal{N}\,\} \\
specS_{connect/3} &= \{\, connect(x, y, d) \mid \text{there exists in } G \text{ a path from } x \text{ to } y \\
&\qquad \text{and } d \text{ is the sequence (list) of its edge labels}\,\} \\
specS_{connect/4} &= \{\, connect(x, y, d, v) \mid connect(x, y, d) \in specS_{connect/3}\,\} \\
specS_{satisfies} &= \{\, satisfies(list, req(maxTrans, maxCost, maxTime)) \mid \\
&\qquad list \text{ is a list of elements of the form } flight(t_i, p_i) \text{ such} \\
&\qquad \text{that the total sum of } p_i\text{'s does not exceed } maxCost, \\
&\qquad p_i, maxCost \in \mathcal{Z}\,\} \\
specS_{analyze} &= \{\, analyze(list, noTrans, totalCost, maxFlightTime) \mid \\
&\qquad list \text{ is a list of elements of the form } flight(t_i, p_i), \\
&\qquad \text{such that the total sum of } p_i\text{'s is equal to } totalCost, \\
&\qquad p_i, totalCost \in \mathcal{Z}\,\} \\
specS_{direct} &= \{\, direct(x, y, flight(t, p)) \mid \text{there exists an edge in } G \\
&\qquad \text{(a direct flight) from } x \text{ to } y \text{ labelled } flight(t, p), \\
&\qquad t, p \in \mathcal{N}\,\} \\
specS_{lesseq} &= \{\, lesseq(X, Y) \mid x \leq y, \quad x, y \in \mathcal{Z}\,\} \\
specS_{add} &= \{\, add(x, y, z) \mid x + y = z, \quad x, y, z \in \mathcal{Z}\,\} \\
specS_{member} &= \{\, member(x, l) \mid l \text{ is a list} \to x \in l\,\}
\end{aligned}
$$

Let us notice that $specS_{direct}, specS_{lesseq}$ and $specS_{add}$ are exact specifications. The remaining ones are approximate: they allow cyclic paths and abstract from the number of transfers and the flight times, and from the form of the last argument of $connect/4$; the approximate specification of $member$ is taken from previous examples.

The second part of specification ($specC$) is constructed from the following sets

of ground atoms:

$specC_{good\_conn}$ = { $good\_conn(x, y, req(k, c, t))$ | there exists in $G$ an
                 acyclic path from $x$ to $y$ denoting a connection such that:
                     the number of transfers does not exceed $k$,
                     its total cost does not exceed $c$,
                     the time of each flight does not exceed $t$,
                 $k, c, t \in \mathcal{N}$ }

$specC_{connect/3}$ = { $connect(x, y, d)$ | there exists in $G$ an acyclic path
                 from $x$ to $y$ such that $d$ is the sequence (list) of its
                 edge labels }

$specC_{connect/4}$ = { $connect(x, y, d, v)$ | there exists in $G$ an acyclic path
                 from $x$ to $y$, $d$ is the sequence (list) of its edge labels
                 and $v$ is a list containing no internal node of the path }

$specC_{satisfies}$ = { $satisfies(list, req(maxTrans, maxCost, maxTime))$ |
                 $list$ is a non-empty list of elements of the form $flight(t_i, p_i)$,
                 the length of $list$ does not exceed $maxTrans + 1$,
                 the total sum of $p_i$'s does not exceed $maxCost$,
                 each $t_i$ does not exceed $maxTime$,
                 $maxTrans, maxCost, maxTime, t_i, p_i \in \mathcal{N}$ }

$specC_{analyze}$ = { $analyze(list, noTrans, totalCost, maxTime)$ |
                 $list$ is a list of elements of the form $flight(t_i, p_i)$ such that:
                     the length of $list$ is equal to $noTrans + 1$,
                     the total sum of $p_i$'s is equal to $totalCost$,
                     each $t_i$ does not exceed $maxTime$,
                 $noTrans \in \mathcal{Z}$, $totalCost, maxTime, t_i, p_i \in \mathcal{N}$ }

$specC_{direct}$ = $specS_{direct}$

$specC_{lesseq}$ = $specS_{lesseq}$

$specC_{add}$ = $specS_{add}$

$specC_{member}$ = { $member(x, l)$ | $l$ is a list $\land\, x \in l$ }

Specifications $specC_{connect/3}$ and $specC_{connect/4}$ are approximate, as they include only acyclic paths while the program also finds paths being concatenation of an acyclic path from $x$ to $y$ with a cycle from $y$ to $y$ (furthermore they require the fourth argument of *connect* to be a list). Specifications $specC_{satisfies}$, $specC_{analyze}$ and $specC_{member}$ are also approximate (as some numbers are required to be in $\mathcal{N}$, the last argument of *analyze* may be not a number, the last argument of *member* not a list, etc.). The remaining specifications above are exact.

Notice that in the description of $specC_{good\_conn}$ the expression "an acyclic path" may be replaced by "a path", as each path with the described properties can be transformed into an acyclic path with these properties by removing the cycles.

Consider Herbrand specification $spec = (specS, specC)$, where

$$\begin{aligned}
specS = \; & specS_{direct} \cup specS_{good\_conn} \cup specS_{connect/3} \cup \\
& specS_{connect/4} \cup specS_{satisfies} \cup specS_{analyze} \cup \\
& specS_{lesseq} \cup specS_{add} \cup specS_{member} \\
specC = \; & specC_{direct} \cup specC_{good\_conn} \cup specC_{connect/3} \cup \\
& specC_{connect/4} \cup specC_{satisfies} \cup specC_{analyze} \cup \\
& specC_{lesseq} \cup specC_{add} \cup specC_{member}
\end{aligned}$$

We would like to prove that FLIGHTS is correct with respect to the above specification *spec*. Thus we have to show that conditions (a) and (b) of Theorem 4.6 are satisfied. To prove condition (a) for predicate *good_conn* we have to show that:

$$specS \models good\_conn(X,Y,R) \leftarrow connect(X,Y,D) \wedge satisfies(D,R)$$

Let $x,y,d,r$ be ground terms such that $specS_{connect/3} \models connect(x,y,d)$ and $specS_{satisfies} \models satisfies(d,r)$. This means that there exists in $G$ a path from $x$ to $y$ and $d$ is the sequence of its edge labels, so $d$ is a non-empty list of elements of the form $flight(t,p)$. Moreover $r$ is of the form $req(k,c,t)$ and the sum of all $p$'s does not exceed $c$. Hence $specS_{good\_conn} \models good\_conn(x,y,r)$.

To prove condition (b) for predicate *good_conn* we have to show that:

$$specC \models good\_conn(X,Y,R) \rightarrow \exists D \; (connect(X,Y,D) \wedge satisfies(D,R))$$

For predicates *connect/3* and *connect/4* we have to prove the following implications:

$$\begin{aligned}
& specS \models connect(X,Y,Dxy) \leftarrow connect(X,Y,Dxy,[X]) \\
& specC \models connect(X,Y,Dxy) \rightarrow connect(X,Y,Dxy,[X]) \\
& specS \models connect(X,Y,[D],V) \leftarrow direct(X,Y,D) \\
& specS \cup specC' \models connect(X,Z,[D|Dyz],V) \leftarrow direct(X,Y,D) \wedge \\
& \qquad\qquad\qquad\qquad \neg member'(Y,V) \wedge connect(Y,Z,Dyz,[Y|V]) \\
& specS \cup specC' \cup spec_= \models connect'(X,Z,L,V) \rightarrow (\exists D \; L = [D] \wedge direct'(X,Z,D)) \\
& \qquad\qquad \vee (\exists Y \; \exists D \; \exists Dyz \;\; L = [D|Dyz] \wedge direct'(X,Y,D) \wedge \\
& \qquad\qquad\qquad \neg member(Y,V) \wedge connect'(Y,Z,Dyz,[Y|V]))
\end{aligned}$$

Let us prove the last implication. Assume that $specC_{connect/4} \models connect(x,z,l,v)$. It means that (in $G$) there exists an acyclic path from $x$ to $z$ and $l$ is the list of its edge labels. If $l$ consists of exactly one edge it must be an edge from $x$ to $z$ (labelled $d$), thus there is a direct flight from $x$ to $z$. So the following holds: $l = [d]$ and $specC \models direct(x,z,d)$. Let $l$ consists of more than one element. So $l = [d_1, d_2, \ldots, d_k]$, where $k > 1$ and $d_i$ is a label of the $i$-th edge of the path from $x$ to $z$. The first edge of that path goes from $x$ to a node $y$, let $d$ be its label ($d = d_1$) and let $dyz = [d_2, \ldots, d_k]$. So the following holds: $l = [d|dyz]$ and $specC \models direct(x,y,d)$. From $specC_{connect/4} \models connect(x,z,l,v)$ it also follows that $v$ is a list such that $y \notin v$, so $specS_{member} \models \neg member(y,v)$. To complete this proof we have to show that $specC \models connect(y,z,dyz,[y|v])$. We already know that there exists an acyclic path from $y$ to $z$ (a subpath of the path from $x$ to $z$), $dyz$ is the list of edge labels of that (sub)path and $v$ is a list such that each internal node of the path from $x$ to $z$ is not its element. It remains to show that also $y$ is not an internal node of the subpath from $y$ to $z$, but this follows immediately from the assumption on the acyclicity of the entire path from $x$ to $y$. This ends the proof of that implication.

We leave to the reader the details of proofs of the remaining implications, as well as proofs of conditions (a) and (b) for the remaining predicates. It follows that FLIGHTS is correct w.r.t. *spec*. Thus we know that if $good\_conn(x, y, req(k, c, t))$ is a computed answer then there exists a connection from $x$ to $y$ and its cost is not greater than $c$. Moreover if a query $good\_conn(x, y, req(k, c, t))$ fails then we know that $specC_{good\_conn} \models \neg good\_conn(x, y, req(k, c, t))$ and there does not exist a connection fulfilling the requirements on the number of transfers, cost and maximal flight time.

From Theorem 4.21 (on proving completeness) it follows, for instance, that if a query $good\_conn(x, Y, req(k, c, t))$ (where $Y$ is a variable) terminates without floundering then all places (vertices) $y$ are found such that there exists a connection from $x$ to $y$ satisfying the requirements (i.e. $specC_{good\_conn} \models good\_conn(x, y, req(k, c, t))$).

Let us extend FLIGHTS by adding a new predicate *bad_conn* defined as follows:

$$bad\_conn(X, Y, Req) \leftarrow \neg good\_conn(X, Y, Req)$$

Let $specn = (specSn, specCn)$, where

$$specSn = specS \cup specS_{bad\_conn} \qquad specCn = specC \cup specC_{bad\_conn}$$

$specS_{bad\_conn}$ = { $bad\_conn(x, y, r)$ | if $r = req(k, c, t)$ and $k, c, t \in \mathcal{N}$ then there does not exist in $G$ a path from $x$ to $y$ denoting a connection such that:
  the number of transfers does not exceed $k$,
  its total cost does not exceed $c$,
  the time of each flight does not exceed $t$ }

$specC_{bad\_conn}$ = { $bad\_conn(x, y, r)$ | there does not exist in $G$ a path from $x$ to $y$ }

Notice that $specS_{bad\_conn}$ and $specC_{bad\_conn}$ also contain atoms of the form $bad\_conn(x, y, r)$ where $x$ or $y$ is not a vertex of the graph. As explained previously, in the description of $specS_{bad\_conn}$ "a path" may be replaced by "an acyclic path". Specification $specS_{bad\_conn}$ is exact and $specC_{bad\_conn}$ is approximate.

We have to prove the following new implications concerning predicate *bad_conn*:

$$specS_{bad\_conn} \cup specC_{good\_conn} \models bad\_conn(X, Y, R) \leftarrow \neg good\_conn(X, Y, R)$$
$$specC_{bad\_conn} \cup specS_{good\_conn} \models bad\_conn(X, Y, R) \rightarrow \neg good\_conn(X, Y, R)$$

As before we leave the details of the proofs to the reader. The rest of implications remains the same, and hence previous proofs remain valid. Thus the extended program is correct w.r.t. the specification *specn*.

So if $bad\_conn(x, y, r)$ is a computed answer then there does not exist a connection from $x$ to $y$ satisfying $r$. After having proved that the program terminates and does not flounder for ground queries of the form $bad\_conn(x, y, r)$, we will also know that the program is complete for such queries w.r.t. *specn*. Hence $bad\_conn(x,y,r)$ will succeed for any $x,y$ such that there does not exist a connection from $x$ to $y$.

To end with, suppose we are interested just in direct connections and choose the following (approximate) specification:

$$specCn_{bad\_conn} = \{\, bad\_conn(x,y,req(0,c,t)) \mid \text{ there does not exist in } G \\ \text{an edge from } x \text{ to } y \text{ labelled } flight(s,p) \text{ such that} \\ p \leq c, \quad c,t,p,s \in \mathcal{N}\,\}$$

This specification refers to the number of transfers (more precisely, to connections without transfers) whereas the specification *specS* abstracted from this. In order to prove correctness of the extended program w.r.t. a specification containing $specCn_{good\_conn}$ instead of $specC_{good\_conn}$, we have to strengthen *specS* (modifying $specS_{good\_conn}, specS_{satisfies}$ and $specS_{analyze}$) and prove again the corresponding conditions. For example a new specification for predicate $good\_conn/3$ could be:

$$specSn_{good\_conn} = \{\, good\_conn(x,y,req(k,c,t)) \mid \text{ there exists in } G \text{ a path} \\ \text{from } x \text{ to } y \text{ of length not greater than } k+1, \text{ the total} \\ \text{cost of the connection does not exceed } c, \quad k,c \in \mathcal{N}\,\}$$

Specifications $specS_{satisfies}$ and $specS_{analyze}$ should be strengthened analogously (by taking into account the total length of the list and the fact that if the costs in the connection list are in $\mathcal{N}$ then the total cost is in $\mathcal{N}$) and the proof should be accordingly modified.

Every (logic) programmer should have, at least in her mind, intended meaning for all the used predicates. Specification $spec = \langle specS, specC \rangle$ is a formalization of such intended meaning. It is important that in most cases the specification is approximate ($specS \neq specC$); specifying exactly the meaning of the program is usually too cumbersome and unnecessary. We believe that the methods advocated in this paper are a formalization of informal reasoning performed by a competent programmer to convince herself about correctness of a program.

## 5 Related work

In this section we present a brief overview of related work.

Due to our approach to specifications, we do not need any explicit notion of precondition, type information, or domain of a procedure. Such notions are used in most other approaches (Bossi and Cocco 1989; Apt 1997; Pedreschi and Ruggieri 1999; Deville 1990) in order to deal with "ill-typed" atoms, for which the behaviour of the program is of no interest. For similar purposes Naish (2000) introduces a 3-valued approach to definite programs.

An approach related to ours is the annotation method of Deransart (Deransart 1993, Section 4; Boye and Małuszyński 1997, Section 4) for proving definite program correctness. It can be seen as refinement of the natural method of Section 3.1, where one proves more (but smaller) implications than those to be proved in the natural method.

A method for proving completeness of definite programs, similar to ours, was presented in (Deransart and Małuszyński 1993). Both approaches are compared in Section 3.3.

Comparison with the operational method (Bossi and Cocco 1989; Apt 1997; Pedreschi and Ruggieri 1999) for correctness of definite programs is given in Section 3.2. The operational method can be generalized to correctness of normal programs (Apt 1995; Pedreschi and Ruggieri 1999); we present a comparison in Section 4.3.4. We show that the correctness proving approaches presented in this paper are stronger than the corresponding operational ones (as far as properties of program answers are concerned); moreover our approach for normal programs is strictly stronger. The method of (Pedreschi and Ruggieri 1999) includes proving completeness of normal programs. In Section 4.4.1 we show that it is strictly weaker than the method of Theorem 4.21. Also, the methods presented in this paper are, in our opinion, simpler than the operational ones. In particular, in the approach of (Pedreschi and Ruggieri 1999) one has to prove correctness, completeness and termination together. Due to this one cannot use approximate specifications.

The comparisons formally show that it is not necessary to refer to operational semantics in reasoning about declarative properties of programs. Naish (1996) presents a similar opinion; he advocates a declarative view for a class of program properties which are often treated as operational.

A related early work is (Deville 1990). It presents a method to develop Prolog programs. Their correctness and completeness follows from construction. However the construction process consists of many stages and is rather complicated.

Our approach to normal programs considers their 3-valued semantics, in contrast to (Deville 1990; Apt 1995; Pedreschi and Ruggieri 1999) where 2-valued semantics is used. The 3-valued completion semantics more precisely corresponds to the operational semantics mainly used in practice (negation as finite failure and SLDNF-resolution). Introducing 3-valued semantics does not result in any difficulties: our proof methods use only the standard 2-valued logic.

An important approach to proving properties of normal programs was proposed by Stärk (1997). It deals with normal programs, executed under Prolog selection rule. Success, failure and termination are described by an inductive theory, called the inductive extension of a program. The theory can be seen as a refinement of the notion of program completion. The program's properties of interest are expressed as formulae and one has to prove that they are consequences of the theory. This is opposite to our approach where properties are expressed as specifications and appear to the left of $\models$, while a program (or a theory similar to program completion) appears to the right of $\models$.

Some properties, like "for any $k$ there exists $l$ such that $P \models p(k, l)$", are expressible in the approach of (Stärk 1997) but cannot be expressed as interpretations in our approach. To deal with such properties we need to use specifications which are theories, not interpretations; we expect that our approach is also applicable to such specifications.

The approach of (Stärk 1997) includes clean termination and is equipped with a tool to mechanically verify the proofs. It is however bound to Prolog selection rule. The involved induction scheme is rather complicated; the scheme seems difficult to use without computer support. The purpose of our work is different: we are interested in the declarative semantics and in basic methods, which can be widely understood by programmers and used – possibly in an informal form – in practical reasoning about actual programs.

Ferrand and Deransart (1993) presented a method of proving correctness of normal programs. (Their terminology is different; what they call "partial completeness" is, in our terminology, correctness for negative atomic queries.) In contrast to our work they deal with the well-founded semantics (van Gelder et al. 1991). Their specifications are thus Herbrand interpretations. The validation conditions of their method consist of conditions equivalent to those of Theorem 4.6, and of additional requirement involving existence of a certain well-founded ordering of atoms.

Malfon (1994) presented methods of proving program completeness for three kinds of semantics, given by the well-founded model, the least fixpoint of $\Phi_P$, and by $\Phi_P \uparrow \omega$. (Notice that the latter is not the Kunen semantics considered in this paper.) Similarly to the previous case, the proposed sufficient conditions for completeness are (equivalent to) the conjunction of our conditions for correctness and a condition involving a well-founded ordering. The latter depends on the considered semantics.

## 6 Conclusions

This paper advocates declarative reasoning about logic programs. We show how to prove correctness and completeness of definite and normal logic programs in a declarative way, independently from any operational semantics. This makes it possible to separate reasoning about "logic" from reasoning about "control". The method for proving correctness of definite programs is not new, however its usefulness has not been appreciated. The methods for completeness and for correctness of normal programs are a contribution of this work.

We refer to two specifications; one for correctness and one for completeness. This makes it possible to specify the program semantics approximately, thus simplifying the specifications and the proofs. In this paper specifications are interpretations, but the approach seems applicable to specifications being theories.

The semantics of normal programs is 3-valued. We do not however explicitly refer to 3-valued logic. Instead, a pair of 2-valued specifications plays a role of a 4-valued specification for correctness and a 3-valued specification for completeness. Also, we use a 2-valued characterization of the 3-valued completion semantics, which may be of separate interest.

Approximate specifications are convenient not only when one deals with program correctness and completeness. We show how approximate specifications can be used to generalize and simplify the proof method of (Apt and Pedreschi 1993) for termination of normal programs.

Some authors suggest referring to operational semantics, in particular to the form of call patterns under LD- (LDNF-) resolution, when reasoning about correctness. We claim that our approach is simpler. We show formally that whatever can be proved using the operational approach, can be proved in the approach advocated here (as far as properties of program answers are concerned). For normal programs our approach is strictly stronger, for definite programs the two approaches are equivalent.

The operational methods additionally prove some properties of the operational semantics, which are outside of the scope of our approach. Obviously, when such properties are of interest, operational methods are indispensable. Their importance should not be neglected. But as long as we are interested in properties of computed answers (correctness and completeness), and not in some details of computations (like call patterns), the declarative approach is sufficient. Termination is an important operational property, which in contrast to correctness and completeness depends on the selection rule. But even for termination most approaches, like the method of (Apt and Pedreschi 1993), do not explicitly refer to call patterns (except for the initial query).

If it were necessary to resort to operational semantics in order to prove basic program properties then logic programming would not deserve to be called a declarative programming paradigm. This work shows that this is not the case.

We believe that the presented proof methods are simple and natural. We claim that they are a formalization of a style of thinking in which a competent logic programmer reasons (or should reason) about her programs. We believe that these methods, possibly treated informally, are a valuable tool for actual everyday reasoning about real programs. We believe that they should be used in teaching logic programming.

**Acknowledgement** The first author was partly supported by the European Commission within the 6th Framework Programme project REWERSE (http://rewerse.net).